\documentclass[journal,twocolumn]{IEEEtran}
\usepackage{cite}
\usepackage{amsmath,amssymb,amsfonts}
\usepackage{algorithmic}
\usepackage{graphicx}
\usepackage{textcomp}
\usepackage{wrapfig}
\usepackage{booktabs}
\usepackage{algorithm} % 您之前加载了 algorithmic 两次
\usepackage{makecell}
\usepackage{stfloats}
\usepackage{url}
\usepackage{hyperref}
\usepackage{array}
\usepackage{siunitx}
\usepackage{subcaption}

\makeatletter
\def\ps@IEEEtitlepagestyle{%
  \def\@oddfoot{\parbox{\textwidth}{\footnotesize\centering
  © 2026 IEEE. Personal use of this material is permitted. Permission from IEEE must be obtained for all other uses, in any current or future media, including reprinting/republishing this material for advertising or promotional purposes, creating new collective works, for resale or redistribution to servers or lists, or reuse of any copyrighted component of this work in other works.\par
  This work has been accepted for publication in IEEE Sensors Journal. DOI: 10.1109/JSEN.2026.3671476}}%
  \def\@evenfoot{\@oddfoot}%
}
\makeatother
\begin{document}
\title{A Passive Software-Defined Radio-based mmWave Sensing System for Blind Integrated Communication and Sensing}
\author{Shiqi Liu, Hang Song, Bo Wei, Nopphon Keerativoranan, and Jun-ichi Takada,%
\thanks{Shiqi Liu is with the Department of Transdisciplinary Science and Engineering, Institute of Science Tokyo, Tokyo 152-8550, Japan. He is also with Okayama University, Okayama 700-8530, Japan.}%
\thanks{Hang Song is with the Research Institute for Semiconductor Engineering, Hiroshima University, Higashi-Hiroshima 739-8527, Japan. He was with the Department of Transdisciplinary Science and Engineering, Institute of Science Tokyo. (e-mail: hanghsong@hiroshima-u.ac.jp)}%
\thanks{Bo Wei is with the Faculty of Environmental, Life, Natural Science and Technology, Okayama University, Okayama 700-8530, Japan, and also with Japan Science and Technology Agency (JST), PRESTO, Kawaguchi, Saitama 332-0012, Japan. (e-mail: weibo@okayama-u.ac.jp)}
\thanks{Nopphon Keerativoranan, and Jun-ichi Takada are with the Department of Transdisciplinary Science and Engineering, Institute of Science Tokyo, Tokyo 152-8550, Japan.}%
}%

% --- 移除 \IEEEtitleabstractindextext 和彩色框 ---
% --- 在 \author{...} 之后立即调用 \maketitle ---

\maketitle

% --- 使用标准的 \begin{abstract} 环境 ---
\begin{abstract}
Integrated Sensing and Communication (ISAC) is considered as a key component of future 6G technologies, especially in the millimeter-wave (mmWave) bands. Recently, the performances of ISAC were experimentally evaluated and demonstrated in various scenarios by developing ISAC systems. These systems generally consist of coherent transmitting (Tx) and receiving (Rx) modules. However, actively transmitting radio waves for experiments is not easy due to regulatory restrictions of radio. Meanwhile, the Tx/Rx should be synchronized and Rx need the information of Tx. In this paper, a fully passive mmWave sensing system is developed with software-defined radio for blind ISAC. It only consists of a passive Rx module which does not depend on the Tx. Since the proposed system is not synchronized with Tx and has no knowledge of the transmitted signals, a differential structure with two oppositely-oriented receivers is introduced to realize the sensing function. This structure can mitigate the influences of unknown source signals and other distortions. With the proposed sensing system, the ambient mmWave communication signals are leveraged for sensing without interrupting the existing systems. It can be deployed for field applications such as signal detection and dynamic human activity recognition since it does not emit signals. The efficacy of the developed system is first verified with a metallic plate with known motion pattern. The measured Doppler spectrogram shows good agreement with the simulation results, demonstrating the correctness of the sensing results. Further, the system is evaluated in complex scenarios, including handwaving, single- and multi-person motion detection. The sensing results successfully reflect the corresponding motions, demonstrating that the proposed sensing system can be utilized for blind ISAC in various applications.
\end{abstract}

% --- 使用标准的 \begin{IEEEkeywords} 环境 ---

% \maketitle % (IEEEtran 类的 \maketitle 必须在 abstract 之前)

\section{Introduction}
\label{sec:introduction}
\IEEEPARstart{I}{ntegrated} Sensing and Communication (ISAC) is recognized as a significant technology for 6G systems by integrating wireless communication and sensing functionalities within shared spectrum and hardware resources~\cite{ref1,ref4,ref5}. In the past decade, ISAC has been widely studied in various aspects including the waveform design \cite{re1}, signal processing technologies \cite{re2,re3}, and channel modeling \cite{re4,re5}. Recently, ISAC has also been investigated in applications such as resource allocation of vehicle networks \cite{re6}, combined with reflective intelligent surface (RIS) technology \cite{re7} and cell-free communication systems \cite{re8,re9}. 

While a lot of ISAC studies are based on theory and simulation, there were some prototype systems developed for evaluating the performance of ISAC \cite{re10,re11}. Among the available ISAC paradigms, communication-centric schemes that exploit communication signals can provide cost-effective solutions and do not interrupt the existing communication structure ~\cite{2024_10openchallange}. Therefore, most of the current ISAC prototype systems were constructed as communication-centric type. Different frequency bands were utilized including sub-6 \cite{re12} and millimeter wave (mmWave) \cite{re11}. The mmWave band is considered to be promising for ISAC applications, as its large bandwidth can facilitate high-resolution localization, and its high carrier frequency can enhance Doppler sensitivity to subtle movements \cite{cui2021integrating, 2024_10openchallange}. In the existing implementations of ISAC prototype, basically the transceiver is built which consists of the transmitting (Tx) module and receiving (Rx) module. And the Tx and Rx modules are synchronized. During the experiment, Tx actively transmits signals, and Rx receives and processes the signals for sensing. In signal processing, the information of Tx is generally necessary. However, this structure may have difficulties in applying to wider scenarios due to the regulatory restrictions for active radio signal emission. 

This paper proposed a fully passive mmWave sensing system with software-defined radio (SDR) for blind ISAC, which is only composed of the Rx module. The passive feature of the proposed system presents several advantages. It avoids the need for synchronization with the Tx, thereby simplifying system design, reducing costs associated with the deployment of dedicated transmitters, and facilitating large-scale deployments. And the full knowledge of Tx is not necessary in the signal processing for sensing. Furthermore, since the proposed passive system does not emit radio waves, it can leverage the existing ambient communication signals and expedite research and development in real-world settings without conflicting the radio regulations. SDR is utilized in developing the passive sensing system for ISAC due to its flexibility and reconfigurability. Meanwhile, it is also promising to be miniaturized with high portability and low-cost by integrated circuits technologies \cite{re14,re15}. SDR has been implemented in many RF sensing systems for various applications \cite{rehman2022development}. The use of commercial-off-the-shelf (COTS) SDR platforms for human motion analysis is an active research area, aiming at developing adaptable, low-cost systems. Moreover, SDRs allow for the integration of multiple receiver channels, which can enhance accuracy by advanced array signal processing techniques. 

In this work, the proposed system utilized a low-cost SDR with two receiving channels on board. Since the system is not coherent with the Tx, the information of transmitted signal is not available. In addition, the local oscillators (LOs) are independent in Tx and Rx, which leads to random and non-deterministic phase relationships between Tx and Rx. The distortions caused by carrier frequency offset (CFO) and phase noise can deteriorate the sensing results \cite{Durr2018}. To solve these problems, this work proposed a differential approach and a signal processing scheme. In the system, two receiver antennas are oppositely oriented. One receiver is configured to capture the quasi-static signal path which is the reference channel. The other receiver is utilized to capture dynamic components affected by object motion which is the sensing channel. Since the two receiving channels share the common LO, by computing the differential channel ratio, the influence of unknown transmitted waveform can be substantially mitigated and common-mode distortions, such as CFO and phase distortion, can be suppressed, thereby extracting motion-induced channel variations. Besides, two mmWave downconverters are utilized to down convert the mmWave signal to the frequency range which can be processed by SDR. To maintain the coherence between the two receiving channels, the two downconverters are also synchronized with a common synthesizer. After receiving the signals, a processing scheme is applied including frame segmentation, alignment, and differential channel calculation. With the proposed sensing system, the ambient mmWave communication signals are leveraged and the differential channel information is utilized for sensing without interrupting the existing systems. 

The proposed system can be deployed for field applications such as signal detection and human activity recognition since it does not emit signals. The system's effectiveness is evaluated through real-world experiments. A calibration measurement is conducted by using a stepper-motor-driven metallic plate with constant speed, and its performance is demonstrated in various scenarios, including single- and dual-person movements.

The main contributions of this work are summarized as follows:
\begin{itemize}
\item A passive mmWave sensing system with differential structure is proposed for blind ISAC, which only consists of Rx module and is independent from Tx.
\item A signal processing scheme is developed for the proposed system, which can mitigate the influences of unknown source signals as well as the distortions caused by CFO and phase noise, extract the dynamic components, and enable the sensing function.
\item The passive mmWave ISAC system is implemented by using COTS SDRs and mmWave downconverters and it is applied to field experiments for motion sensing. The efficacy of the proposed system is demonstrated by the experiment results.
\end{itemize}

The reminder of the paper is organized as follows. In section II, a brief review of the existing sensing system, especially ISAC, is given. Section III presents the detailed system design of the proposed passive mmWave ISAC system. In Section IV, the signal processing scheme is depicted for realizing the sensing through the received signals. Section V shows the experimental setup for performance verification and Secion VI demonstrates the experiment results. Finally, the conclusion is made in Section VII. 

\section{Related Work}

There are several works which developed the prototype systems for ISAC. These systems work at different frequencies, utilize different waveforms, and are applied to various purposes. They are generally based on SDR. ISAC systems can be categorized into two types, the active sensing and passive sensing \cite{Zhang2022JCASurvey, chetty2}. In active sensing, the Tx and Rx are typically co-located and the synchronization is established. The information of Tx is fully known to the Rx. The Tx and Rx are highly cooperative in active sensing. While in passive sensing, the Tx and Rx are spatially separated and the synchronization is generally not achieved. According to the knowledge level of the Tx signal, the passive sensing can be further characterized into two kinds. In some passive ISAC systems, the Rx has the knowledge of the protocol and signal structure, and can conduct channel estimation. Here, this kind of system is denoted as semi-cooperative. In other passive ISAC system, the Tx and Rx are totally non-cooperative, where the Rx captures the signals from the illuminator of opportunity and no prior information of the transmitted signal is available. A comparison of the existing ISAC systems is summarized in Table~\ref{tab:landscape}.

\begin{table*}[htbp]
\centering
\caption{ {Comparison of ISAC systems.}}
\label{tab:landscape}

{
\begin{tabular}{|c|c|c|c|c|c|}
\hline
\textbf{Related Work} & \textbf{Category} & \textbf{Signal} & \textbf{Frequency} & \textbf{Tx-Rx} & \makecell{\textbf{Differential} \\ \textbf{Structure}} \\ \hline
\cite{chetty1, chetty2} & Passive & WiFi & 2.4 GHz & Non-Cooperative & Yes \\ \hline
\cite{Li2018} & Passive & ISM Band & 915 MHz & Non-Cooperative & Yes \\ \hline
\cite{Chu2025} & Passive & Digital Television & 530 MHz & Non-Cooperative & Yes \\ \hline
\cite{Zhang2022} & Passive & WiFi & 5.825 GHz & Semi-Cooperative & No \\ \hline
\cite{Li2017} & Passive & WiFi & 5 GHz & Semi-Cooperative & No \\ \hline
\cite{re12} & Passive & OFDM & 900 MHz, 2.45 GHz & Semi-Cooperative & No \\ \hline
\cite{re23} & Passive & OFDM, Chirp & 5.9 GHz & Semi-Cooperative & No \\ \hline
\cite{re29} & Passive & OFDM & 2.4 GHz & Semi-Cooperative & No \\ \hline
\cite{re11, re16, re17} & Active & OFDM & 24 GHz, 76--81 GHz & Cooperative & No \\ \hline
\cite{re21, re22, re28} & Active & OFDM & 28 GHz & Cooperative & No \\ \hline
\cite{re10, re25} & Active & OFDM & 60 GHz & Cooperative & No \\ \hline
\cite{re27} & Active & Chirp & 26 GHz, 71--76 GHz & Cooperative & No \\ \hline
\cite{re30} & Active & FMCW radar & 79 GHz & Cooperative & No \\ \hline
\textbf{This work} & Passive & mmWave Communication & 25.1 GHz & Non-Cooperative & Yes \\ \hline
\end{tabular}
}
\end{table*}

In non-cooperative passive ISAC systems, the signals of opportunity are utilized  and the receiver exploits two receiving channels as a differential structure for sensing. Chetty et al. utilized WiFi signals at 2.4 GHz for human activity and respiration sensing \cite{chetty1, chetty2}. Li et al. developed a passive sensing system by utilizing a 915 MHz ISM-band signal for health monitoring applications including body movements, respiration, and physical activities \cite{Li2018}. Chu et al. utilized digital television signals at 530 MHz to detect low--slow--small targets \cite{Chu2025}. 

In semi-cooperative passive ISAC systems, channel state information (CSI) is widely utilized for sensing, such as WiFi CSI. Zhang et al. proposed a cross-domain gesture recognition with WiFi by developing a general model to fit various conditions \cite{Zhang2022}. Li et al. proposed an indoor human tracking system by commodity WiFi \cite{Li2017}. 
Apart from existing communication signals, there are also works which utilized other signal waveforms to implement ISAC. Khan et al. designed an SDR-based platform for activity recognition by using orthogonal frequency division multiplexing (OFDM) signals, which operates in both 900 MHz and 2.45 GHz \cite{re12}. Moro et al. investigated the potential by using ISAC for unmanned aerial vehicle (UAV) synthetic aperture radar (SAR) imaging \cite{re23}. The proposal operated in 5.9 GHz. The simulation was based on OFDM communication signal and the pulse compression of OFDM signal was carried out. SAR imaging was carried out with time domain back projection method. In the experimental implementation, the SDR was utilized and chirp signal was transmitted. Xu et al. designed an OFDM-based MIMO ISAC testbed operating at 2.4 GHz by using SDR \cite{re29}. The performance of the dual-functional waveform design was validated. 

On the other hand, in mmWave frequency bands, the research on passive ISAC system utilizing existing communication signal is limited. Many existing ISAC systems are active type and the OFDM signal is utilized for evaluation.  Ozkaptan et al. developed a software defined OFDM radar for ISAC in the 76-81 GHz spectrum \cite{re16}. It was utilized for the automotive scenario, and the simultaneous video streaming and range-Doppler sensing were demonstrated. The same group also proposed a 24 GHz mmWave multiple-input and multiple-output (MIMO) ISAC testbed and the performance in multi-target detection was verified \cite{re17,re11}. The OFDM transceiver was designed and the CSI variations were exploited for sensing with machine learning techniques. Guan et al. proposed a 3-D imaging system by using the 28-GHz OFDM 5G-like communication signals \cite{re21,re22}. The signal processing pipeline was developed by leveraging mathematical similarities between OFDM and frequency modulated continuous wave (FMCW). Yang et al. proposed a mmWave ISAC prototype system operating at 28 GHz which can realize multiple functions \cite{re28}. The OFDM signal frame is designed in accordance with 5G standards. Maletic et al. developed a real-time ISAC system which can support high-data rate \cite{re10, re25}. This system operates at 60 GHz and the OFDM signal is utilized. Pham et al. presented two mmWave ISAC testbeds at both 26 GHz and 71-76 GHz \cite{re27}. The FMCW waveforms (chirps) were utilized in the measurement. Cui et al. developed mmRipple towards ISAC, which utilized the mmWave FMCW radar to enable communication function via smartphone vibrations \cite{re30}.

As shown above, there are many passive ISAC systems operating in frequency bands lower than several GHz. However, in the mmWave band, most ISAC studies are active sensing which rely on cooperative configurations. In contrast, the proposed mmWave ISAC system is passive and non-cooperative, enabling blind sensing by directly leveraging existing ambient mmWave communication signals.

\section{System Design}

\begin{figure*}[!t]
    \centering
    \includegraphics[width=0.95\textwidth]{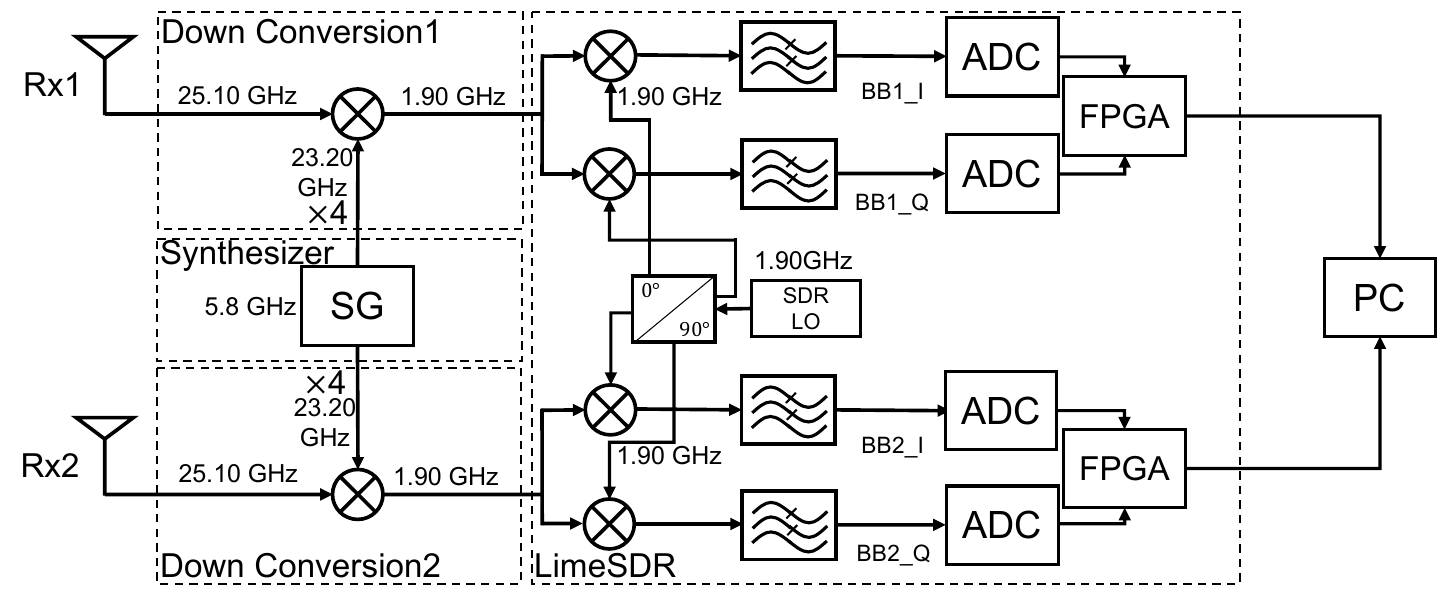}
    \caption{Schematic diagram of the proposed passive mmWave ISAC sensing system.}
    \label{fig:system}
\end{figure*}

The schematic diagram of the proposed sensing system is shown in Fig.~\ref{fig:system}, which is based on a differential Rx structure. The system design and operation principle are detailed in this section. 

\subsection{Differential Receiver Architecture}

Since the sensing system does not rely on the Tx, the differential Rx architecture is proposed to mitigate the influence of the transmitted signal as well as the distortions caused by CFO and phase noise. As shown in Fig.~\ref{fig:system}, two Rx channels on the same SDR are utilized. As the SDR basically works in several GHz, the mmWave-range downconverters are cascaded to enable the acquisition of mmWave signals. Two directional mmWave antennas, designated as Rx1 and Rx2, are connected to the two separate downconverters, respectively. These antennas receive the mmWave signals and feed them to the system. A key aspect of the proposed system is involving a dual-channel differential architecture, designed to coherently capture and process ambient signals for passive sensing. Thus, a single synthesizer provides a common LO signal to both downconverter modules on the two receiving channels. After down conversion, the intermediate frequency (IF) signals are routed to SDR. The SDR uses the common internal LO and sampling clock for both channels, thereby a fully coherent dual-receiver system is established. After being down-converted to the complex baseband, the I/Q baseband signals are digitized for signal processing to realize sensing. 

\begin{figure}[t]
    \centering
    \includegraphics[width=\linewidth]{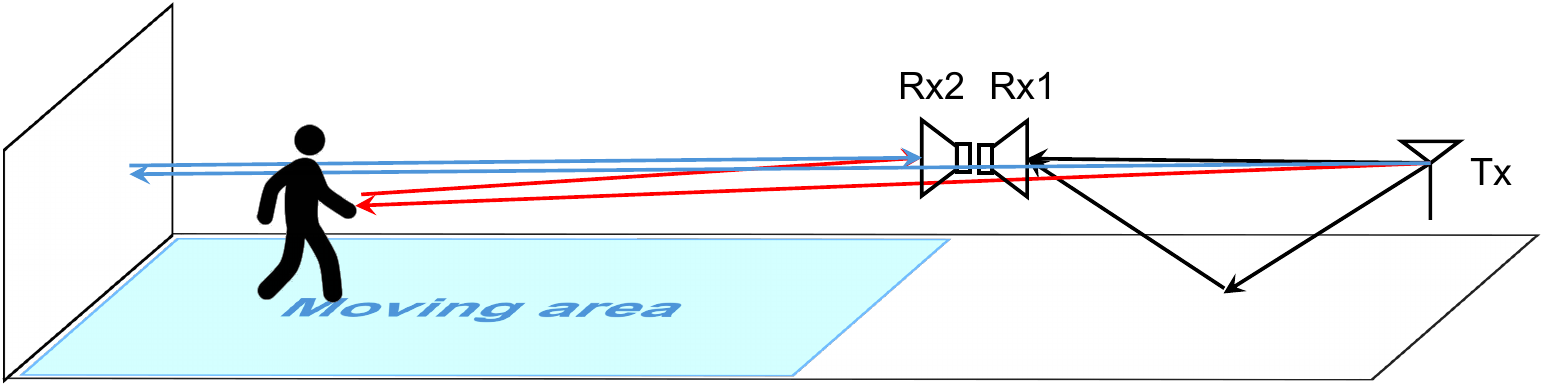}
    \caption{Deployment geometry of the differential architecture for passive ISAC sensing. (Black lines: line-of-sight (LOS) and other static MPC to Rx1. Red lines: dynamic component from moving target to Rx2. Blue lines: other static MPCs such as reflection from wall to Rx2.)}
    \label{fig:placement}
\end{figure}

The deployment geometry of the proposed differential system is essential for the sensing. A conceptual graph of the deployment geometry is depicted in Fig.~\ref{fig:placement}. The two Rx antennas are placed in opposing orientations. Rx1 is designated as the reference antenna. It is oriented towards the Tx to capture the line-of-sight (LOS) path from the Tx as well as other static multipath reflections. This establishes a stable, quasi-static reference channel. On the other hand, Rx2 is designated as the sensing antenna. It is directed towards the region of interest to sense the object motions. This configuration enables Rx2 to primarily capture dynamic signal components reflected from the moving object as well as other static multipath components (MPC) such as reflection from the wall, while being less sensitive to the direct path from the source. This oppositional geometry maximizes the differential channel variation when motion occurs, facilitating the extraction of time-varying characteristics without requiring knowledge of the transmitted waveform.

\subsection{Formulation of the Sensing Principle}

Since the proposed system is not synchronized with the Tx, the distortions such as CFO and phase noises exist. Considering that the human motion is subtle which only induces maximum several-hundred frequency shift, the influence of the distortions are assumed to be non-negligible, which can overwhelm the effect caused by the object motion and make the motion signature undetectable. In addition, the lack of the Tx signal information makes it difficult to extract accurate motion-induced features from the received signals. Here, the proposed differential architecture is able to overcome these problems. This subsection details the formulation of the sensing principle for passive mmWave blind ISAC with the proposed system.

Regarding the deployment geometry as shown in Fig.~\ref{fig:placement}, the channel responses for both receiving channels are modeled. Denote the channel response of the reference channel as $H_{\text{1}}(f, t)$ which is considered to be quasi-static. It can be expressed as:
\begin{align}
    H_{\text{1}}(f,t) &= \sum_{n=1}^{N} g_1(\theta_n)A_ne^{-j2\pi f \tau_n} \label{eq:Hs1}
\end{align}
where $n$ is the index of the MPC in reference channel. \textit{N} is the number of the MPCs. $g_{\text{1}}(\theta_n)$ is the gain of antenna Rx1 as a function of the angle of arrival $\theta_n$. $A_n$ and $\tau_n$ are the complex amplitude and propagation delay of the \textit{n}th component, respectively. The complex amplitude $A_n$ accounts for various factors including path loss and phase shifts from reflections. Similarly, denote the channel response of the sensing channel as $H_{\text{2}}(f,t)$, which is considered. $H_{\text{2}}(f,t)$ is composed by the static and dynamic components. Denote the static components as $H_{\text{s2}}(f)$. Denote the dynamic components as $H_{\text{dyn}}(f,t)$ which is time-variant and influenced by the target motion. Then, $H_{\text{2}}(f,t)$ can be expressed as:

\begin{align}
    H_{\text{2}}(f,t) &= H_{\text{s2}}(f) + H_{\text{dyn}}(f,t) \nonumber \\ % <-- 关键：在这里也加上 \nonumber
    &= \sum_{m=0}^{M} g_2(\theta_m)A_me^{-j2\pi f \tau_m}  \\ %\label{eq:Hs2_Hdyn}
    & \quad + \sum_{l=0}^{L} g_2(\theta_l)A_l(t) e^{-j2\pi f (d_l(t)/c)}  \nonumber
\end{align}
where $c$ is the speed of light, $m$ is the index of the static MPC in sensing channel. \textit{M} is the number of the static MPCs. $g_{\text{2}}(\theta_m)$ is the gain of antenna Rx2 as a function of the angle of arrival $\theta_m$. $A_m$ and $\tau_m$ are the amplitude and propagation delay of the \textit{m}th static component, respectively. $l$ is the index of the dynamic MPC in sensing channel. \textit{L} is the number of the dynamic MPCs. $g_{\text{2}}(\theta_l)$ is the gain of antenna Rx2 as a function of the angle of arrival $\theta_l$. $A_l(t)$ and $d_l(t)$ are the amplitude and path distance change of the \textit{l}th dynamic component, respectively. $d_l(t)$ is considered to be caused by the target motion and consequently introduces the Doppler frequency shift in the received signal. $A_l(t)$ is also modeled as time-variant due to the path change.

Denote the Tx signal as $S(f,t)$ in the frequency domain, which is assumed to change over time and unknown. Then, the received signals at Rx1 and Rx2 antennas are defined as $R_{\text{1}}(f,t)$ and $R_{\text{2}}(f,t)$ in frequency domain, respectively. Here, a sufficiently high signal-to-noise ratio (SNR) is assumed, allowing the effects of additive white gaussian noise (AWGN) to be disregarded. Therefore, the received signals can be expressed as:
\begin{align}
    R_{\text{1}}(f,t) &\approx S(f,t) \cdot H_{\text{1}}(f,t) \cdot e^{j(2\pi\Delta f_{\text{off}}t +\Delta\phi(t))} \cdot H_{\text{dev}} \label{eq:R1_ideal} \\
    R_{\text{2}}(f,t) &\approx S(f,t) \cdot H_{\text{2}}(f,t) \cdot e^{j(2\pi\Delta f_{\text{off}}t +\Delta\phi(t))}\cdot H_{\text{dev}}\label{eq:R2_ideal}
\end{align}
where $\Delta f_{\text{off}}$ is the CFO in the system and $\Delta\phi(t)$ is the random time-variant phase noise. Since both channels share common LOs, these distortion terms are identical. $H_{\text{dev}}$ is the channel variation induced by the hardware system. This term is also considered common to both channels as the two receive chains work in exact the same manner within the single dual-channel SDR \cite{lms7002m-wiki}. If there is only one receiving channel, the motion-related Doppler frequency cannot be correctly estimated due to the CFO and other distortions. On the other hand, by computing the ratio of the two received signals owing to the differential architecture, the influence of CFO, the unknown transmitted signal term and all common-mode distortion terms are canceled. Therefore, the motion-induced variation can be better extracted for sensing. Then, a relative channel response can be obtained as:
\begin{equation}
    H_{\text{rel}}(f,t) = \frac{R_{\text{2}}(f,t)}{R_{\text{1}}(f,t)} =  \frac{H_{\text{s2}}(f)}{H_{\text{1}}(f,t)} + \frac{H_{\text{dyn}}(f,t)}{H_{\text{1}}(f,t)} \label{eq:H_rel_final}
\end{equation}

As demonstrated above, the differential operation cancels the unknown transmitted signal and all common-mode components, including CFO, phase noise, and the device-induced channel variation. The resulting relative channel response $H_{\text{rel}}(f,t)$ isolates the environmental characteristics, where the static component $H_{\text{s2}}(f)/H_{\text{1}}(f,t)$ can be subsequently removed by subtracting the temporal mean, leaving only the motion-induced dynamic component. Then, to obtain a robust time-varying characteristics of the dynamic component from this relative channel response, a spectral averaging is applied in the subsequent signal processing. The details are explained in Section IV-C. By processing $H_{\text{rel}}(f,t)$, the object motion feature can be extracted and the sensing is realized.

\section{Signal Processing Scheme}

Following signal acquisition, the raw in-phase (I) and quadrature (Q) samples from both receiver channels are processed through a multi-stage procedure in order to extract motion-induced Doppler signatures. The processing scheme is designed to handle the irregular, non-cooperative feature of the ambient signals. The details are presented in this subsection including: (A) preprocessing and system calibration, (B) frame segmentation and alignment, (C) differential channel calculation, and (D) Doppler Spectrogram estimation. In this section, since the acquired signals are sampled in time domain, the discrete representation is utilized.

\subsection{Preprocessing and System Calibration}
As an initial step, the raw I/Q data streams acquired from the SDR are calibrated to correct for inherent hardware-related impairments, including DC offsets and I/Q imbalance. The DC offsets are removed by subtracting the running mean from the time-domain signals. Then, the I/Q imbalance is compensated by utilizing the built-in calibration function of the LimeSuite API, which corrects for gain and phase mismatches between the in-phase and quadrature paths. After these preprocessing steps, the calibrated signals are utilized for subsequent processing stages.

\subsection{Frame Segmentation and Alignment}
\begin{figure}[ht]
\centering
\includegraphics[width=\linewidth]{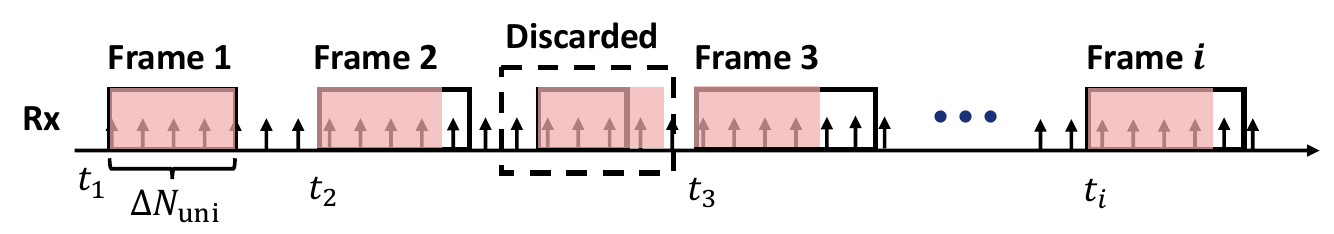}
\caption{Illustration of aligned frames used for analysis. Pink part is the uniform window size.}
\label{fig:frame_normalization}
\end{figure}

Since the information of transmitted signals is not known, the received signals should be segmented to recognize different signal frames. Here, the ambient signals are assumed to be received with irregular duration and period, as illustrated in Fig.~\ref{fig:frame_normalization}. Through sampling, the received signal from reference channel after calibration is defined in discrete form as $r_{\text{1}}[k]$. $k$ is the discrete-time index that specifies the sampling instant. It is a positive integer. Similarly, the received signal from sensing channel after calibration is defined as $r_{\text{2}}[k]$. To identify the frames, an energy-based detection algorithm is proposed, which is formalized in Algorithm~\ref{alg:segmentation}. The purpose is to identify the start time and the duration of each identified signal frame. The frame segmentation procedure is divided into three stages: (1) rising-edge detection to identify the start time of a signal frame, (2) frame boundary detection to determine the end time of the signal frame, and (3) saving the start time and duration of each frame.

Two thresholds are utilized to determine the start time and the end time of the frame, which are denoted as $A_{\text{th,st}}$ and $A_{\text{th,ed}}$, respectively. During the process, the amplitude of $r_{\text{1}}[k]$ is first calculated. Then, the amplitude of each signal sample is compared with $A_{\text{th,st}}$ one by one along with time. When the amplitude is larger than $A_{\text{th,st}}$ for the first time, the time is recorded and saved into a list $S_{\text{st}}$. Then, the signal amplitude is compared with the other threshold $A_{\text{th,ed}}$. When the amplitude is smaller than $A_{\text{th,ed}}$ for the first time, the duration of the signal frame is calculated by subtracting the start time from the end time. And the duration is also saved to another list $W_{\text{dur}}$. With the start time and duration, one signal frame can be identified. Next, the comparison with $A_{\text{th,st}}$ is repeated to identify the subsequent frames until the end of $r_{\text{1}}[k]$. Finally, the start times and durations for all the frames can be retrieved from $S_{\text{starts}}$ and $W_{\text{dur}}$.

\begin{algorithm}[hbt!]
\caption{Frame Segmentation Algorithm}
\label{alg:segmentation}
\begin{algorithmic}[1]
\STATE \textbf{Input:} Calibrated reference signal $r_{\text{1}}[k]$, start threshold $A_{\text{th,st}}$, end threshold $A_{\text{th,ed}}$, continuous end count $C_{\text{end}}$.
\STATE \textbf{Output:} Frame start time list $S_{\text{st}}$, Frame duration list $W_{\text{dur}}$.

\STATE Initialize $S_{\text{st}} \leftarrow \emptyset$, $W_{\text{dur}} \leftarrow \emptyset$.
\STATE Initialize $\text{in\_signal} \leftarrow \text{false}$, $\text{end\_counter} \leftarrow 0$.

\STATE Compute amplitude $A[k] \leftarrow |r_{\text{1}}[k]|$.

\FOR{$idx=1$ to length of $A[k]$}
    \IF{$\text{in\_signal}$}
        \IF{$A(idx) < A_{\text{th,ed}}$}
            \STATE $\text{end\_counter} \leftarrow \text{end\_counter} + 1$.
        \ELSE
            \STATE $\text{end\_counter} \leftarrow 0$.
        \ENDIF
        
        \IF{$\text{end\_counter} \geq C_{\text{ed}}$}
            \STATE $\text{end\_idx} \leftarrow idx - C_{\text{ed}} + 1$.
            \STATE Add $(\text{end\_idx} - \text{start\_idx})$ to $W_{\text{dur}}$.
            \STATE $\text{in\_signal} \leftarrow \text{false}$.
        \ENDIF
    \ELSE
        \IF{$A(idx) > A_{\text{th,st}}$}
            \STATE $\text{start\_idx} \leftarrow idx$.
            \STATE Add $\text{start\_idx}$ to $S_{\text{st}}$.
            \STATE $\text{in\_signal} \leftarrow \text{true}$.
            \STATE $\text{end\_counter} \leftarrow 0$.
        \ENDIF
    \ENDIF
\ENDFOR
\STATE \textbf{return} $S_{\text{st}}$, $W_{\text{dur}}$.
\end{algorithmic}
\end{algorithm}

Since subsequent signal processing is generally performed in frequency domain, a statistical approach is proposed to determine a uniform frame size, which ensures consistent dimensions for FFT processing. Here, the $x_p$ percentile of all frame durations from $W_{\text{dur}}$ is calculated and chosen as the standard window size, denoted as $\Delta N_\text{uni}$. Then, all detected frames are aligned by trimming and only the initial $\Delta N_\text{uni}$ samples are retained. Note that when the frame size is smaller than $\Delta N_\text{uni}$, this frame is discarded and will no longer be used. This process results in a set of uniform-sized frames, as illustrated in Fig.~\ref{fig:frame_normalization}. The pink part shows the chosen window size and the signal within this portion is denoted as
\begin{align}
    r_{1,t_i}[k] = r_{\text{1}}[k], k \in [S_{\text{st}}[i], S_{\text{st}}[i]+\Delta N_\text{uni}-1]  
     \label{eq:r[k]_frame}
\end{align}
where $i$ is the index of the aligned signal frame. $t_i$ is the start time of the $i$th frame. Define the sampling interval as $T_{\text{s}}$. Then, $t_i$ equals $S_{\text{st}}[i]\cdot T_{\text{s}}$. These aligned frames are utilized for all the following processes. The frame between Frame 2 and Frame 3 is shorter than the window size, thus it is discarded. 

\subsection{Differential Channel Calculation}
This stage processes the aligned frames to generate a dynamic-only time series of the relative channel response. For each $i$th aligned frame, a $\Delta N_\text{uni}$-point fast Fourier transform (FFT) is applied to $r_{1,t_i}[k]$ and $r_{2,t_i}[k]$, estimating its frequency-domain representation defined in Eqs.~\eqref{eq:R1_ideal} -- \eqref{eq:R2_ideal} as $R_{\text{1}}(f_q,t_i)$ and $R_{\text{2}}(f_q,t_i)$, respectively. Here $f_q = q\Delta f$ is a discrete frequency sample at $q$ frequency index with a frequency resolution of $\Delta f = 1/(\Delta N_\text{uni} T_s)$. Note that the time intervals between two adjacent signal frames are not necessarily identical.

\begin{figure}[ht]
     \centering
     \begin{subfigure}[b]{0.48\textwidth}
     \centering
    \includegraphics[width=\textwidth]{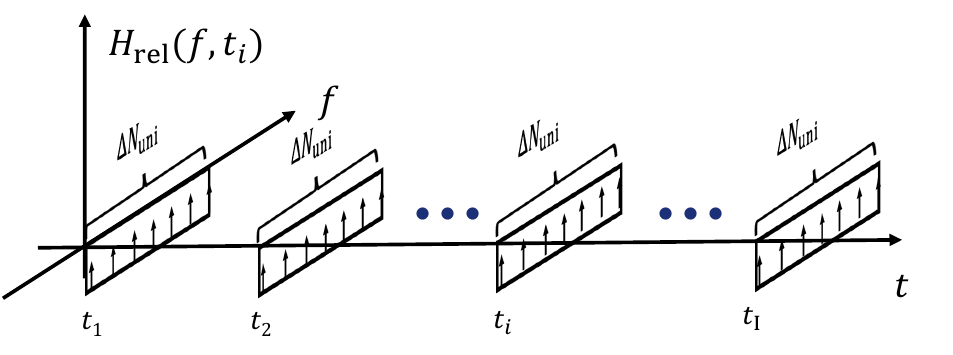}
    \caption{Concept diagram}
    \label{fig:concept}
    \end{subfigure}
    \hfill
    \begin{subfigure}[b]{0.48\textwidth}
    \centering
    \includegraphics[width=\textwidth]{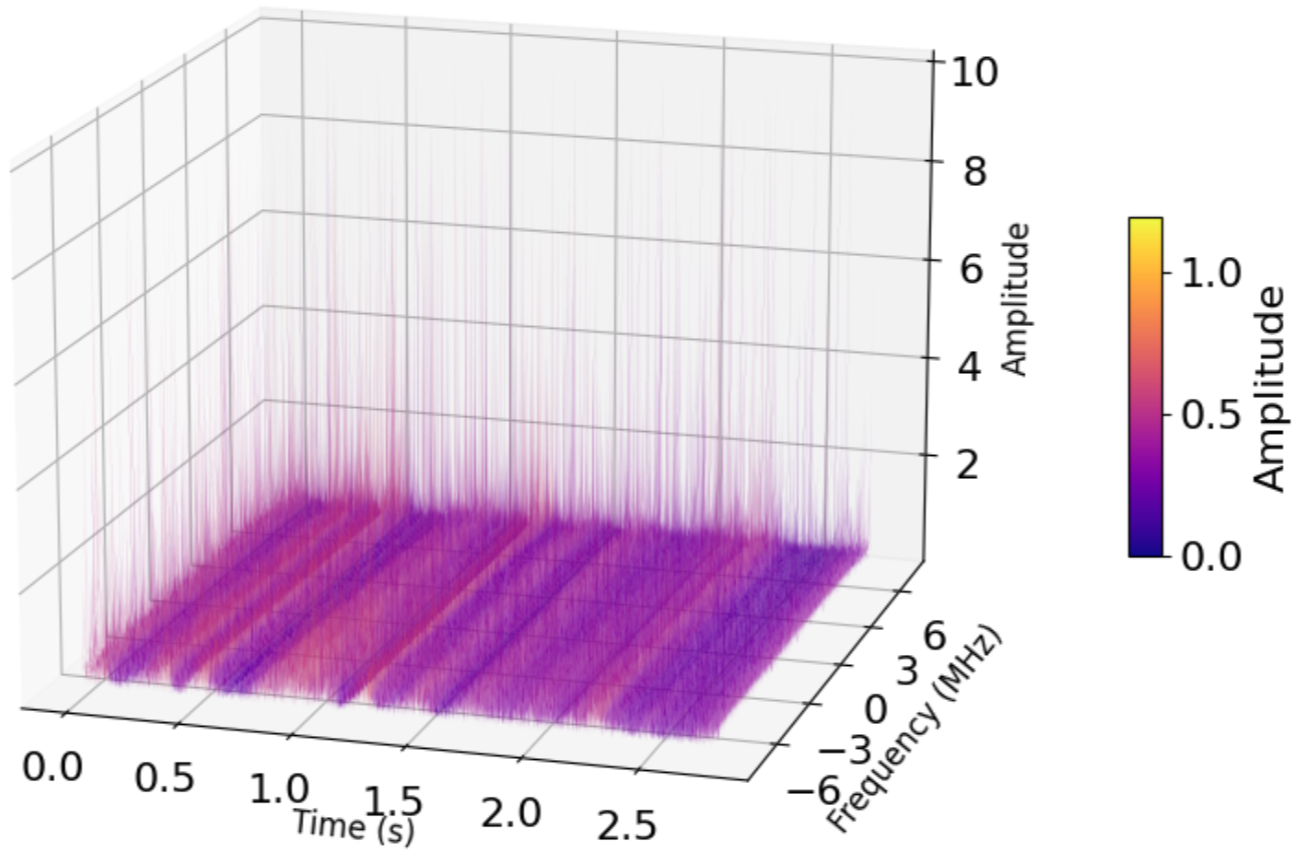}
    \caption{Measurement result}
    \label{fig:reality}
    \end{subfigure}
\caption{Relative channel response $H_\text{rel}(f_q, t_i)$ in time-frequency domain. (a) Concept diagram. (b) Measurement result.}
    \label{fig:channel_concept_vs_reality}
\end{figure}

The relative channel response $H_\text{rel}(f_q, t_i)$ is obtained from $R_{1}(f_q, t_i)$ and $R_{2}(f_q, t_i)$ using Eq. \eqref{eq:H_rel_final}. Since the FFT is utilized, the number of bins in the frequency domain is also $\Delta N_\text{uni}$ as illustrated in Fig.~\ref{fig:channel_concept_vs_reality}(a). However, this operation is susceptible to numerical instability. When the magnitude of the reference channel $|R_{1}(f_q, t_i)|$ approaches zero in certain frequency bins (i.e. a deep spectral null), the division may result in large, spike-like artifacts. One example is shown in Fig.~\ref{fig:channel_concept_vs_reality}(b), which is from the measurement data.

To get a better estimation of the channel response, these artifacts are mitigated through a spectral averaging step. This process is based on the assumption that the signal bandwidth is smaller than the channel coherence bandwidth, allowing the channel to be approximated as a flat fading model. To justify this averaging, the relative channel response can be approximated with a simplified model as:
\begin{equation}
    H_\text{rel}(f_q,t_i) = \frac{H_2(f_q,t_i)}{H_1(f_q,t_i)} = \frac{A_{2} e^{-j2\pi(f_c+f_q)\tau_2(t_i)}}{A_{1} e^{-j2\pi(f_c+f_q)\tau_1(t_i)}}
    \label{eq:spectral_ratio_plat}
\end{equation}
where $A_{1}$ and $A_{2}$ are the amplitudes, and $\tau_1(t_i)$ and $\tau_2(t_i)$ are the propagation delays. $f_c$ is the center frequency. When summing up Eq. \eqref{eq:spectral_ratio_plat} over the symmetric index $q$, the phase components related to $f_q$ tend to be canceled out. Consequently, the frequency-dependent response can be merged into a single robust equivalence at the center frequency. In the signal processing scheme, this spectral averaging is performed as:
\begin{equation}
    H_{\text{rel}}(t_i) = \frac{1}{\Delta N_\text{uni}} \sum_{q=-\frac{\Delta N_\text{uni}}{2}+1}^{\frac{\Delta N_\text{uni}}{2}} H_\text{rel}(f_q, t_i)
    \label{eq:spectral_avg_expanded}
\end{equation}

This process transforms the noisy frequency-dependent response $H_{\text{rel}}(f_q, t_i)$ into a robust frequency-independent equivalence $H_{\text{rel}}(t_i)$. Compared to using a single-bin value as the channel response, leveraging the entire spectrum provides a better estimation, effectively suppressing the spike-like artifacts.

Finally, to isolate motion-induced variations from the relative channel response, the temporal mean of entire time series $\{H_{\text{rel}}(t_i)\}$ is computed and subtracted from each $H_{\text{rel}}(t_i)$ as:
\begin{equation}
     \tilde{H}_{\text{rel}}(t_i) = H_{\text{rel}}(t_i) - \text{Mean}(\{H_{\text{rel}}(t_i)\})
     \label{eq:temporalMean}
\end{equation}
This operation removes the static baseline, and the resulting signal $\tilde{H}_{\text{rel}}(t_i)$ is the dynamic-only time series of the relative channel response used for Doppler spectrum estimation.

Algorithm~\ref{alg:estimation} summarizes the procedure for the entire differential channel calculation described above.

\begin{algorithm}[H]
\caption{Differential Channel Calculation}
\label{alg:estimation}
\begin{algorithmic}[1]
\STATE \textbf{Input:} Aligned signal frames $r_{1,t_i}[k]$, $r_{2,t_i}[k]$; Frame size $\Delta N_\text{uni}$.
\STATE \textbf{Output:} Dynamic-only time series $\tilde{H}_{\text{rel}}(t_i)$.

\FOR{each frame $i$}
    \STATE $R_{1}(f_q, t_i) \leftarrow \text{FFT}(r_{1,t_i}[k])$.
    \STATE $R_{2}(f_q, t_i) \leftarrow \text{FFT}(r_{2,t_i}[k])$.
    \STATE $H_{\text{rel}}(f_q, t_i) \leftarrow R_{2}(f_q, t_i) / R_{1}(f_q, t_i)$.
    \STATE $H_{\text{rel}}(t_i) \leftarrow \text{Mean}_{f}(H_{\text{rel}}(f_q, t_i))$.
\ENDFOR

\STATE $\tilde{H}_{\text{rel}}(t_i) \leftarrow H_{\text{rel}}(t_i) - \text{Mean}(\{H_{\text{rel}}(t_i)\})$ .
\STATE \textbf{return} $\tilde{H}_{\text{rel}}(t_i)$.
\end{algorithmic}
\end{algorithm}

\subsection{Doppler Spectrogram Estimation}
The final stage of the signal processing is to extract the Doppler signatures from $\tilde{H}_{\text{rel}}(t_i)$. Because the intervals of two adjacent frames are generally non-uniform, the Non-Uniform Short-Time Fourier Transform (NU-STFT) is employed to compute the time-varying Doppler spectrogram as:
\begin{equation}
S(\tau, f_d) = \sum_{i} \tilde{H}_{\text{rel}}(t_i) \cdot w(t_i - \tau) \cdot e^{-j2\pi f_d t_i}
\label{eq:nustft}
\end{equation}
where $w(\cdot)$ is a windowing function centered at time $\tau$, and $f_d$ represents the Doppler frequency. By investigating the estimated Doppler spectrogram, the motion signatures can be revealed for subsequent analysis.

\section{Experimental Setup}
To assess the proposed passive sensing system and characterize its performance, a series of experiments were conducted within a typical indoor office environment with the size of 5.5~m $\times$ 7.8~m.

\subsection{System Configuration and Deployment}
The hardware platform, as detailed in Section~III, was utilized for all measurements. A photograph of the system in operation is shown in Fig.~\ref{fig:hardware_closeup}. In this study, the system is configured to be operated at 25.1~GHz for performing real-world experiment. A license-free commercial transmitter (NTG-2501) operating at 25.1~GHz was used as the ambient signal source, which is not cooperative with the sensing system. The downconverters utilized in this work are ADMV1014 (Analog Devices) which covers a wide frequency range from approximately 24~GHz to 44~GHz, allowing for flexible adaptation to various operational environments. The synthesizer is ADF4372 (Analog Devices). The LO frequency generated from the synthesizer is 5.8~GHz and input to the downconverter. Each ADMV1014 multiplies the shared LO by a factor of four to generate a \SI{23.2}{\giga\hertz} LO signal for mixer. By mixing with \SI{23.2}{\giga\hertz} LO signal, the ambient \SI{25.10}{\giga\hertz} signals are coherently down-converted to IF of \SI{1.9}{\giga\hertz} for both channels. Subsequently, the IF signals are routed to the single dual-channel SDR (LimeSDR-USB, Lime Microsystems) via the high-frequency input ports.

In the system deployment, the distance between the Tx and the Rx1 was set to approximately 2.2~m. A baseline separation of 1~cm was maintained between the Rx1 and Rx2, with the antennas placed in a back-to-back configuration. All antennas were positioned at a uniform height of 1~m above the floor. The SDR sampling rate was set to 15~MS/s. The $x_p$,  which determines the window size of signal frame, is set to 5th percentile. {The key system parameters are summarized in Table~\ref{tab:system_parameters}.

\begin{table}[htbp]
    \caption{{Key System Parameters}}
    \label{tab:system_parameters}
    \centering
    {
    \begin{tabular}{l c}
        \toprule
        \textbf{Parameter} & \textbf{Value} \\
        \midrule
        Transmitter Model & JRC NTG-2501 \\
        Receiver Model &LimeSDR-USB \\
        Tx-Rx module & Non-cooperative \\
        Antenna Model & Pasternack PEWAN028-10KF \\
        Antenna Gain & 10 dBi \\
        Downconverter & Analog Devices ADMV1014 \\
        Synthesizer & Analog Devices ADF4372 \\
        Operating Frequency & 25.1 GHz \\
        LO Frequency & 5.8 GHz (Synthesizer) \\
                     & 23.2 GHz (Downconverter) \\
        IF Frequency & 1.9 GHz \\
        ADC Sampling Rate & 15 MS/s \\
        Tx-to-Rx1 Distance & 2.2 m \\
        Antenna Height & 1.0 m \\
        \bottomrule
    \end{tabular}
   }
    
\end{table}

\subsection{Measurement Scenarios}
Two sets of experiments were designed to validate the accuracy of the system and evaluate the performance in realistic sensing scenarios. One is the controlled scenario using a metallic plate. The other is a set of human activity scenarios.

\subsubsection{Controlled Validation Scenarios}
In this set of experiments, the measurements were conducted by using a square metallic plate which was mounted on a linear slider equipped with a stepping motor as shown in Fig.~\ref{fig:motor_setup_photo}. The size of the plate is $10~\text{cm} \times 10~\text{cm}$. During the experiment, the plate was moved with a controlled constant speed by programming the stepping motor. The initial distance from the sensing antenna (Rx2) to the plate was 3.3~m, ensuring the plate remained in the far-field region. With this configuration, the system's Doppler measurement accuracy can be quantitatively verified. Two motion patterns were designed for the validation. One is that the plate was programmed to move towards and away from the receiver in turn at a constant speed of 3.125~cm/s for a duration of 3.2~s. The other is that the plate was programmed to move continuously towards the receiver at a constant velocity. 

\begin{figure}[ht]
    \centering
    \subfloat[]{\includegraphics[width=0.98\linewidth]{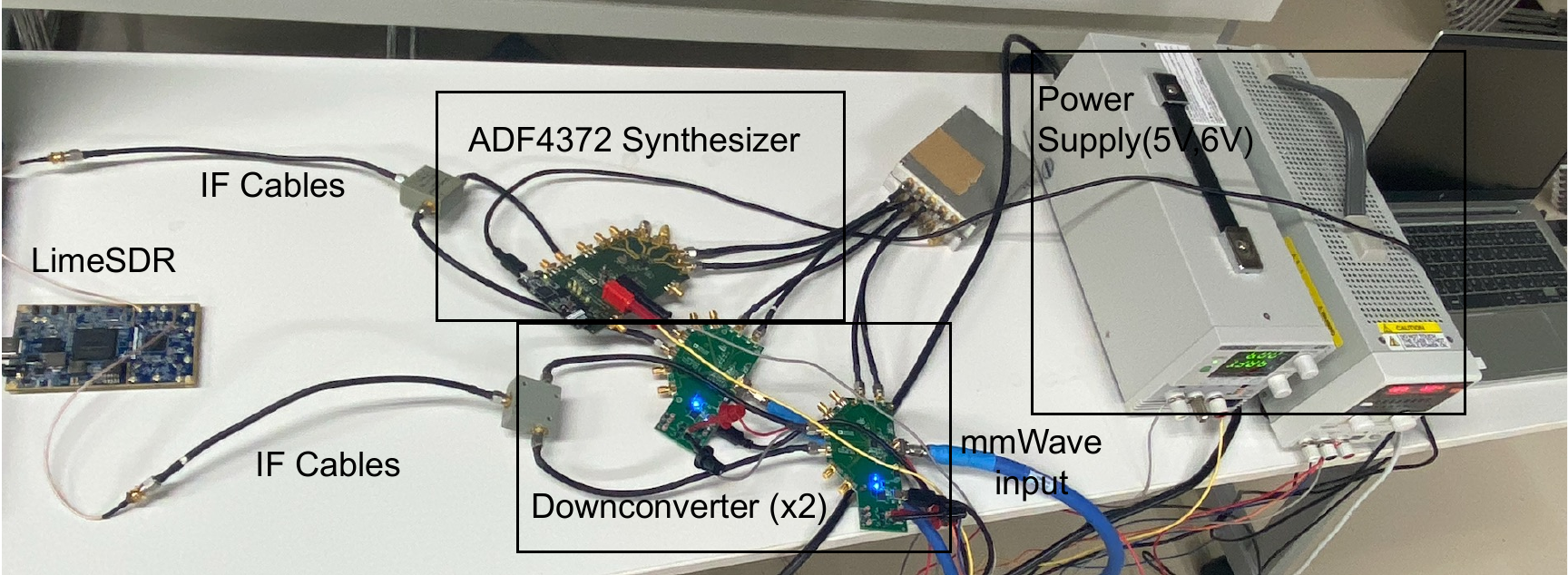}\label{fig:hardware_closeup}}
    \\
    \subfloat[]{\includegraphics[width=0.98\linewidth]{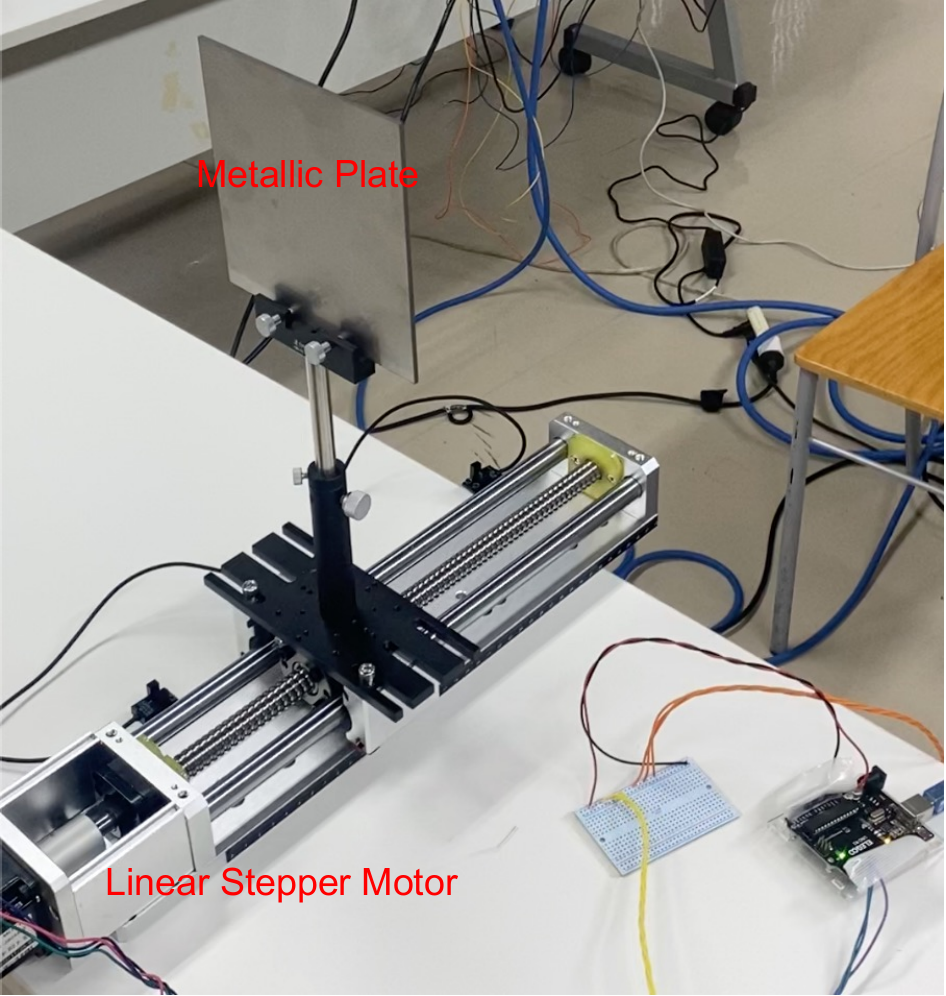}\label{fig:motor_setup_photo}}
    \caption{Experimental hardware and validation setup. (a) A close-up view of the proposed system in operation. (b) The controlled validation setup, featuring a metallic plate mounted on a linear slider equipped with a stepping motor.}
    \label{fig:combined_setup}
\end{figure}

\subsubsection{Human Activity Scenarios}
Following the quantitative system validation, a broad set of representative human activities was performed within the deployment area depicted in Fig.~\ref{fig:human_deployment}. The scenarios included: a static background, hand wave, unidirectional walk, back-and-forth walk, and two-person walk. In static background scenarios, data was collected with no human presented in the sensing area to establish a baseline of the static case. In hand wave scenario, a human waved hand in proximity to the sensing system. In unidirectional walk scenarios, a human walked towards the sensing system and stopped. In the back-and-forth walk scenario, a human walked towards and then far away from the sensing system repeatedly. In two-person walk scenarios, one human walked back-and-forth in a relatively far position. Another human walked near the sensing system.

\begin{figure}[ht]
    \centering
    \includegraphics[width=0.95\linewidth]{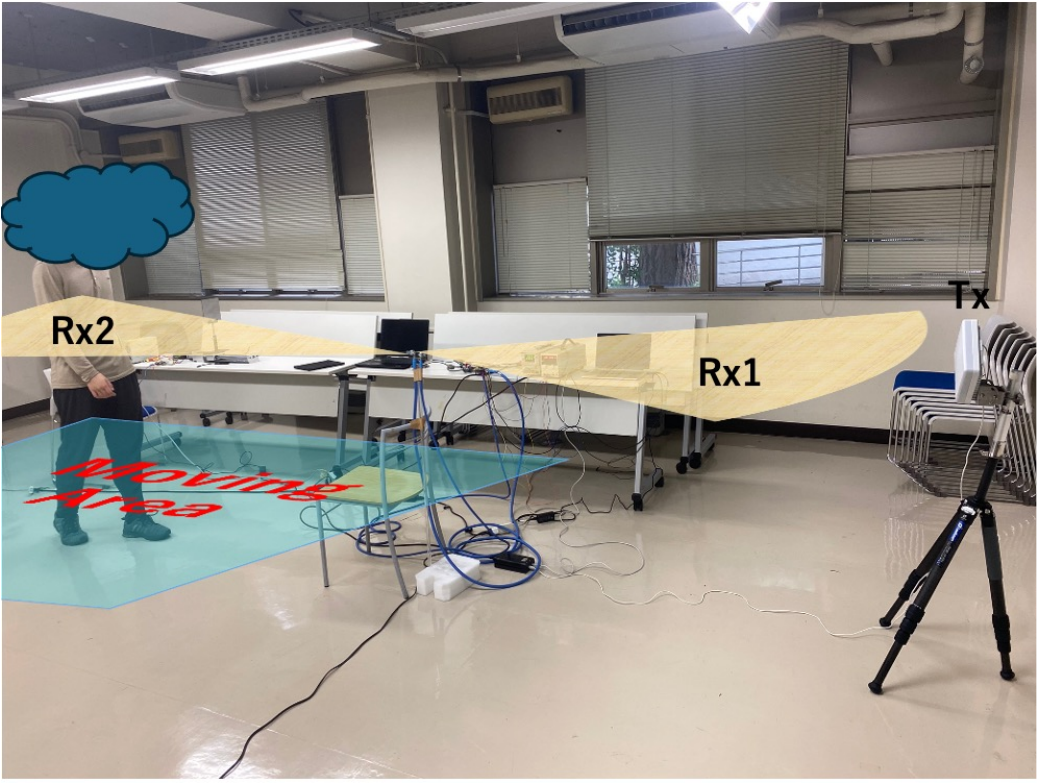}
    \caption{Deployment for human activity sensing, illustrating the physical layout of the transmitter (Tx), the ISAC sensing system, and the designated area for human movement.}
    \label{fig:human_deployment}
\end{figure}

\section{Experiment Results}
This section presents the experimental results in the two sets of scenarios, including the validation in controlled condition and human activity sensing. 

\subsection{System Validation and Calibration}
To validate the efficacy and accuracy of the proposed system, a theoretical ground truth is first established via simulation for the controlled plate moving scenario. Then, the measurement result is compared with the simulation and a timing calibration factor is derived to correct the spectrogram estimation. Finally, the calibration factor is verified by another measurement.

\subsubsection{Simulation Methodology and Ground Truth}
A theoretical baseline for the controlled experiments was generated using physical optics (PO) approximation method~\cite{Balanis2012AEE}. The scattering model of plane wave from a flat rectangular plate is utilized. In the validation step, the Tx and Rx were set to be perpendicular to the metal plate. Therefore, zenith angles of the incident and reflected wave are close to 0. In such condition, the total scattered electric field can be expressed as:
\begin{equation}
  E^{s} =  -jE_0\frac{ab\beta}{2\pi}\,\frac{e^{-j\beta r}}{r} \label{eq:PO_E_total}
\end{equation}
where $a$ and $b$ are the sizes of the plate. $\beta$ is the wavenumber. $E_0$ is the amplitude of the electric field. $r$ is the distance from plate to Rx2.

Using Eq.\eqref{eq:PO_E_total}, the simulated received signal $E^{s}(t)$ is generated at each time step $t$ based on the plate's trajectory. The Doppler spectrogram is then computed from $E^s(t)$. When the plate moves forward and backward at the speed of 3.125~cm/s for a duration of 3.2 s for each direction, the theoretical Doppler spectrogram is shown in Fig.~\ref{fig:sim_pulsed_motion}. The Doppler frequency shifts are $\pm$5.23~Hz in towards and backwards directions, respectively. These results are utilized as the ground truth for performance validation.

\begin{figure}[ht]
    \centering
    \includegraphics[width=0.9\linewidth]{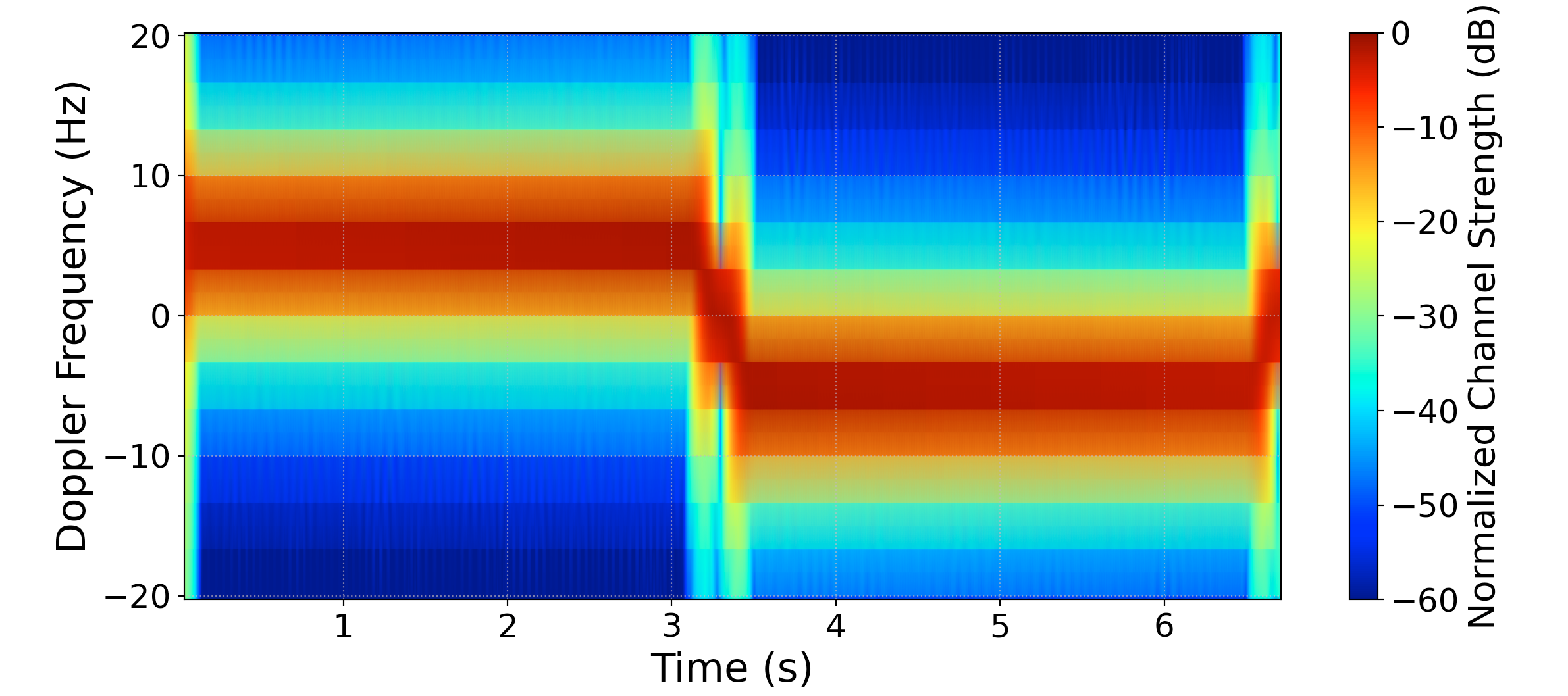}
    \caption{Doppler spectrogram generated from the simulated signals.}
    \label{fig:sim_pulsed_motion}
\end{figure}

\subsubsection{Measurement and System Calibration}
The Doppler spectrogram generated from the measurement data is shown in Fig.~\ref{fig:exp_pulsed_motion}(a). A notable discrepancy is observed can be noticed that the peak Doppler frequency is around $\pm$8.9 Hz compared with $\pm$5.23 Hz in simulation as shown in Fig.~\ref{fig:sim_pulsed_motion}. It is also observed that the motion duration is only displayed as 1.9 seconds although the real duration is 3.2 seconds in measurement. This temporal compression is caused by the data acquisition and transferring loop. During the measurement, the raw I/Q data are acquired by SDR and transferred to host PC repeatedly as data block in real time. Since the data transfer and saving also consumes some time, the data during the transfer period is not acquired. Therefore, the actual saved data duration is shorter than the real operation time. To correctly estimated the Doppler spectrogram, the real time interval for the received signals are needed. Consequently, a linear time-scaling factor $k_t$ is derived by:
\begin{equation}
    k_t = \frac{\text{Theoretical Duration}}{\text{Measured Duration}} = \frac{3.2\,\text{s}}{1.9\,\text{s}} \approx 1.7
\end{equation}
Applying this factor to $t_i$, the calibrated start time of each signal frame $t_{i,\text{cal}}$ can be calculated as:  
\begin{equation}
t_{i,\text{cal}} = k_t \cdot t_i \label{Eq:time_cal}
\end{equation}
Using the calibrated time, the Doppler spectrogram is corrected as shown in Fig.~\ref{fig:exp_pulsed_motion}(b). It can be observed that the motion duration and peak Doppler shift now show good agreement with the simulated one.

\begin{figure}[ht]
    \centering
    \subfloat[]{\includegraphics[width=0.98\linewidth]{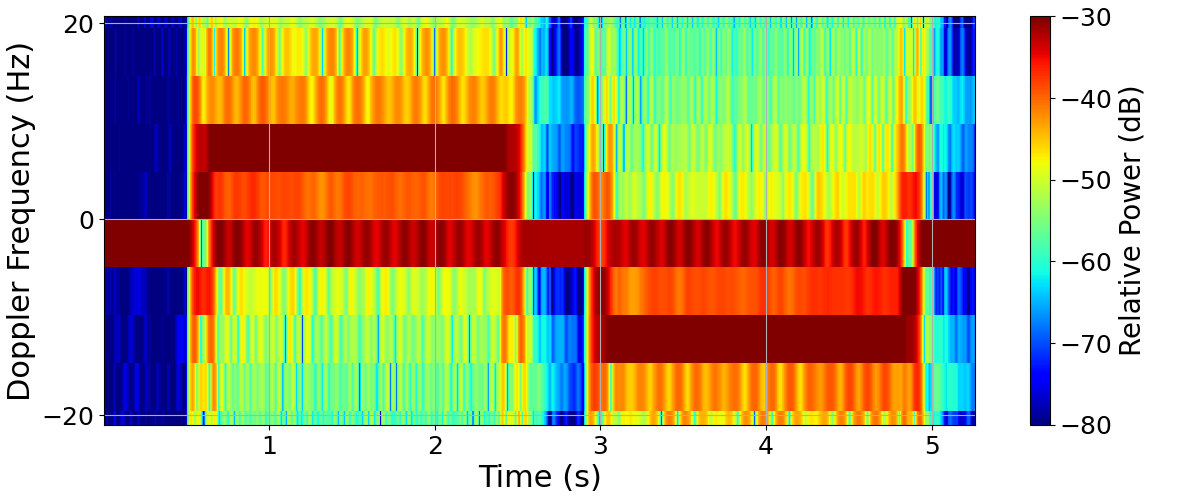}\label{fig:uncalibrated_exp}}
    \\
    \subfloat[]{\includegraphics[width=0.98\linewidth]{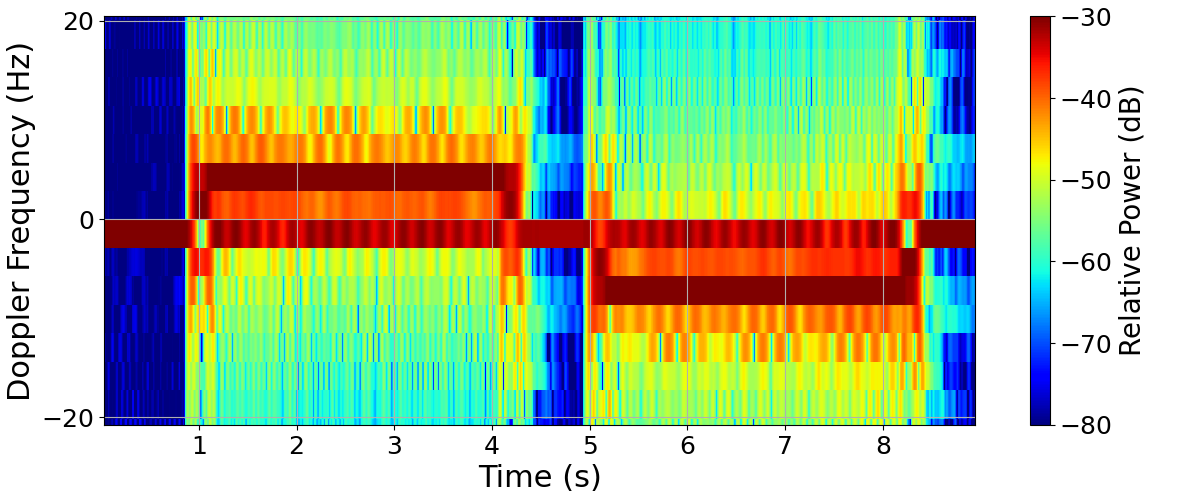}\label{fig:calibrated_exp}}
    \caption{Doppler spectrograms generated from measurement data using (a) uncalibrated time $t_i$ and (b) calibrated time $t_{i,\text{cal}}$.}
    \label{fig:exp_pulsed_motion}
\end{figure}

\subsubsection{Calibration Factor Verification}
To verify the generality of time-scaling factor $k_t$, it was applied to the data obtained from another continuous plate motion experiment.

The scenario with a constant speed of 3.125~cm/s was tested. The Doppler spectrogram calculated with $t_{i,\text{cal}}$ from measurement is shown in Fig.~\ref{fig:cont_spectrogram}. It displays a stable Doppler signature throughout the entire movement period. For comparison, the Doppler spectrum was computed by using signal $\tilde{H}_{\text{rel}}(t_{i,\text{cal}})$ over the whole duration. The result is depicted in Fig.~\ref{fig:cont_global_doppler_fast}. A distinct peak is observed at +5.23~Hz, which quantitatively matches the theoretical value. The notch at 0~Hz is the consequence of the temporal mean subtraction process depicted in Eq. \eqref{eq:temporalMean}, where the DC component of $\tilde{H}_{\text{rel}}(t_{i,\text{cal}})$ is removed. This process is conducted because the static components are not of interest in the Doppler analysis. Figure~\ref{fig:cont_spectrogram} also reveals periodical subtle modulations in the spectrogram. This is considered to be caused by plate micro-vibrations during the movement driven by stepping motor, indicating the system's high sensitivity.

\begin{figure}[ht]
    \centering
    \subfloat[]{\includegraphics[width=0.98\linewidth]{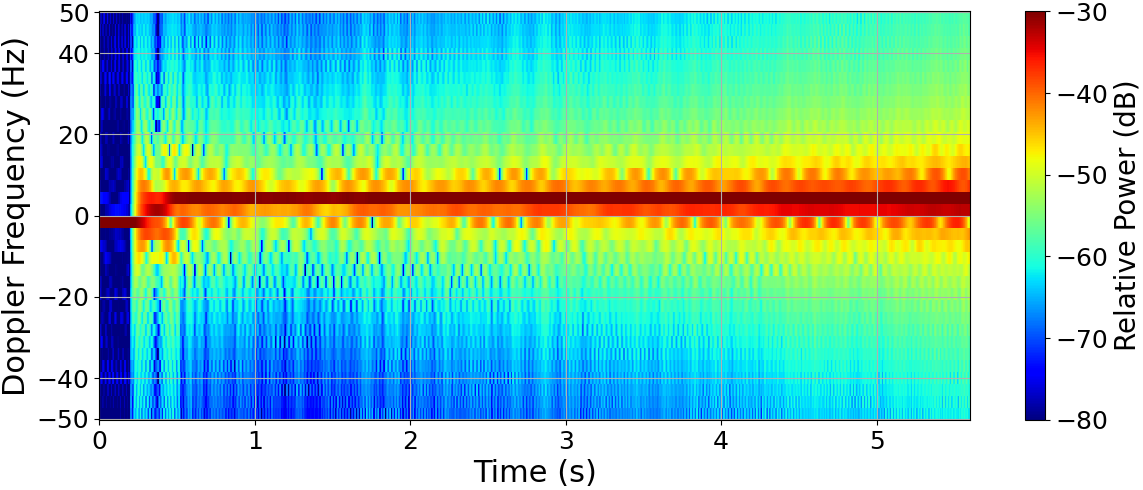}\label{fig:cont_spectrogram}}
    \\ 
    \subfloat[]{\includegraphics[width=0.98\linewidth]{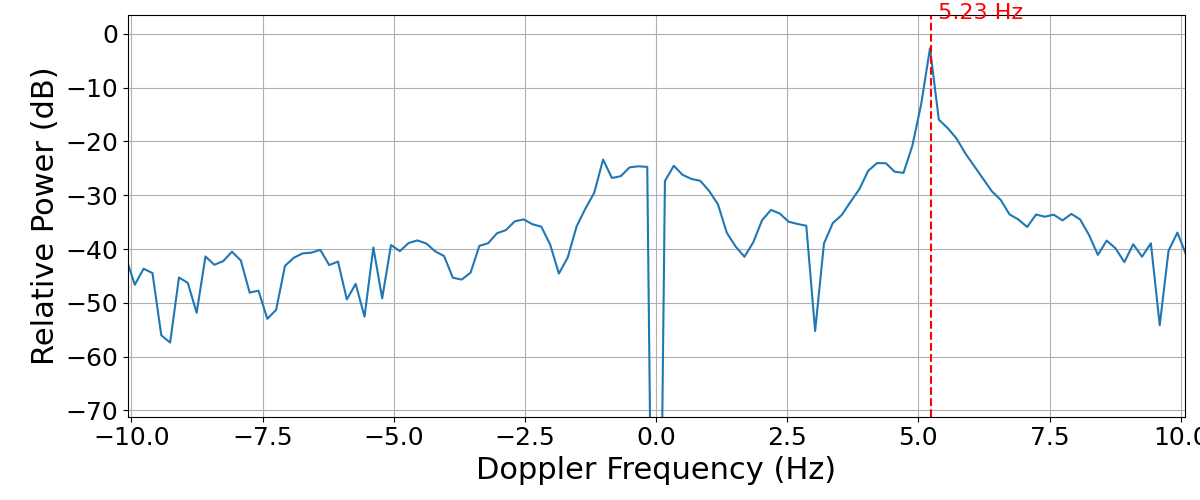}\label{fig:cont_global_doppler_fast}}
    \caption{Experiment results from the continuous plate motion scenario. (a) Doppler spectrogram and (b) Doppler spectrum over the whole measurement duration.}
    \label{fig:validation_continuous_fast}
\end{figure}
\subsection{Human Activity Sensing Results}
Following the validation of the system's accuracy, a series of experiments were conducted to evaluate its performance across scenarios of increasing complexity. The purpose was to investigate the system's capability to capture and differentiate the Doppler signatures generated by various human movements. Totally five scenarios were considered as depicted in Section V.A(2). All spectrogram results were generated using the calibrated time $t_{i,\text{cal}}$ as derived in Section V-A.

\begin{figure*}[!t]
\centering
\subfloat[No Human (Static Background)]{\includegraphics[width=0.48\linewidth]{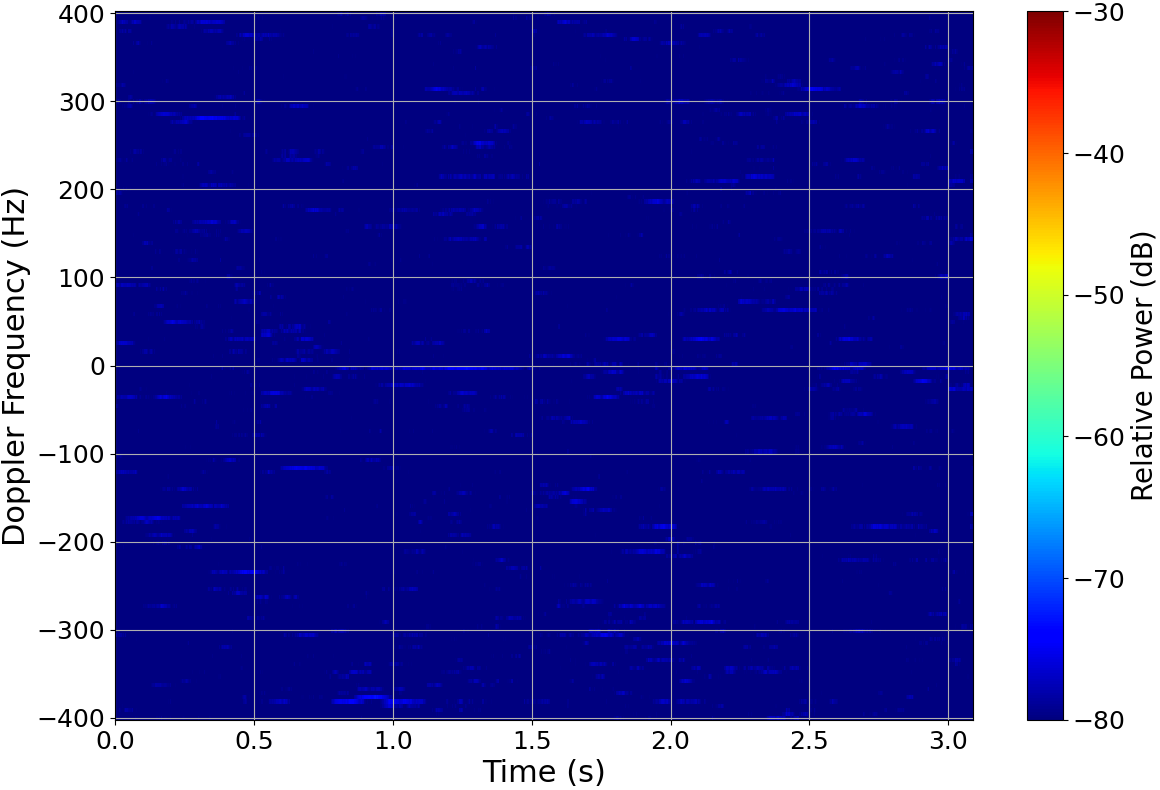}\label{fig:no_human}}
\hfill
\subfloat[Unidirectional Walking]{\includegraphics[width=0.48\linewidth]{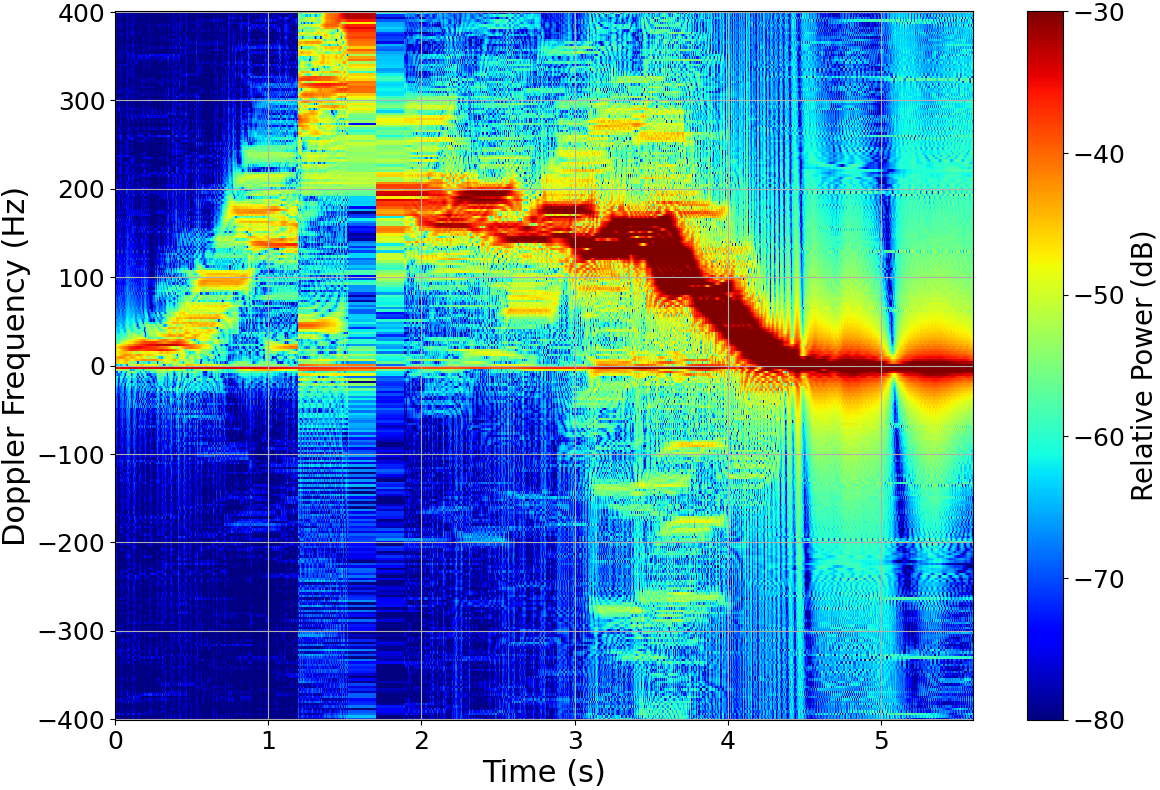}\label{fig:unidir_walk}}
\caption{Doppler Spectrograms under (a) static background scenario and (b) unidirectional walking scenario with one human subject.}
\label{fig:baseline_simple}
\end{figure*}

\begin{figure*}[!t]
\centering
\subfloat[Hand Wave]{\includegraphics[width=0.48\linewidth]{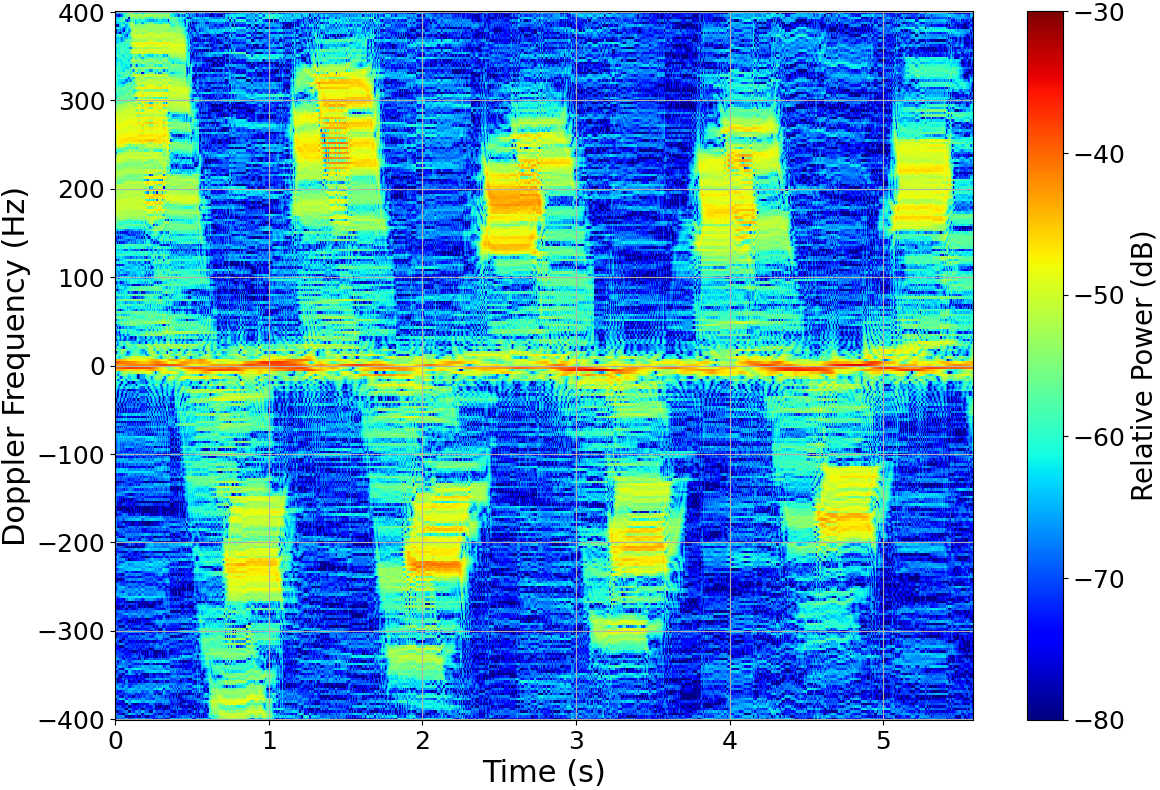}\label{fig:hand_wave}}
\hfill
\subfloat[Back-and-Forth Walking (1 Person)]{\includegraphics[width=0.48\linewidth]{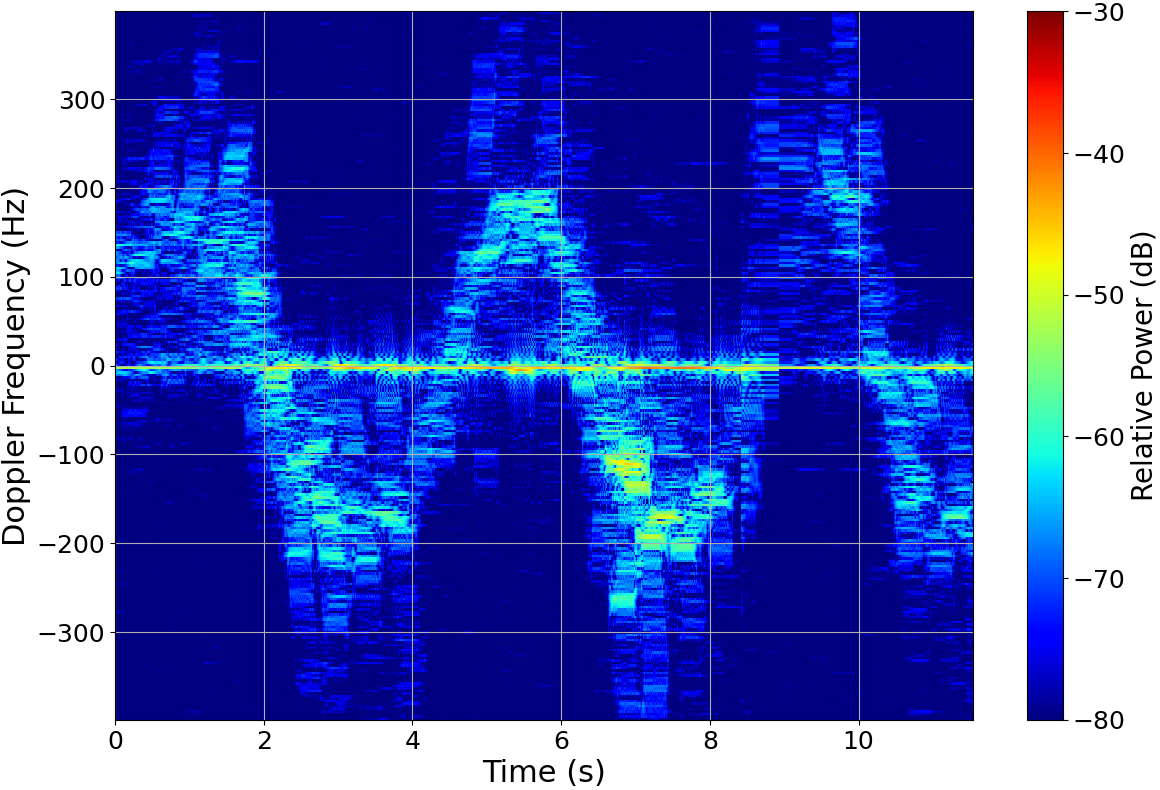}\label{fig:one_person_walk}}
\caption{Doppler spectrograms under periodic motion scenarios. (a) Hand-waving of one human. (b) Back-and-forth walking of one human }
\label{fig:periodic}
\end{figure*}

\subsubsection{Static and Unidirectional Motion}
The system's performance was first evaluated in a static background scenario where there was no human in the sensing area. The Doppler spectrogram is shown in Fig.~\ref{fig:baseline_simple}(a). As anticipated, the spectrogram exhibits no recognizable Doppler signature. The the channel strength remains below -80~dB. This result demonstrates that the differential sensing architecture and signal processing scheme effectively suppress both hardware-related noise and static environmental clutter.

Then, the fundamental capability to detect motion was evaluated under a unidirectional walking scenario. In this scenario, the human subject walked towards the sensing system and stopped. The Doppler spectrogram is in Fig.~\ref{fig:baseline_simple}(b). A continuous positive Doppler signature is observed with a maximum frequency shift of approximately +200~Hz, corresponding to the subject walking towards the sensing antenna. It is also observed that the channel strength of the reflection increases significantly from around -55dB $\sim$ -20dB as the human subject approaches the receiver. The successful capture of these motion characteristics confirms the sensing system's capability to detect a moving target.

\subsubsection{Periodic Human Motions}
The system's capacity to distinguish between different types of activities was investigated by analyzing periodic movements. First, a human waved hands periodically near the sensing system. The Doppler spectrogram from this scenario is shown in Fig.~\ref{fig:periodic}(a). It can be observed that the hand-waving gesture is recognized by a signature of rapid alternating positive and negative Doppler shifts reaching up to $\pm$350~Hz.

Then a scenario where one human subject walks back and forth was evaluated. The Doppler spectrogram from this scenario is presented in Fig.~\ref{fig:periodic}(b). A detailed examination of its spectrogram reveals a rich structure containing distinct Doppler components. The main part of the Doppler signature is considered to be related to the torso's movement, whose frequency peak is approximately $\pm$200~Hz. Furthermore, there are some weaker spectral components extending to $\pm$250~Hz and beyond which can be observed. These higher frequency components are considered to be the characteristic of human gait and likely generated by the faster-moving limbs during the gait cycle. The ability to resolve various features within a single motion implies the system's potential for fine-grained human activity analysis.

\begin{figure}[!t]
\centering
\includegraphics[width=0.95\linewidth]{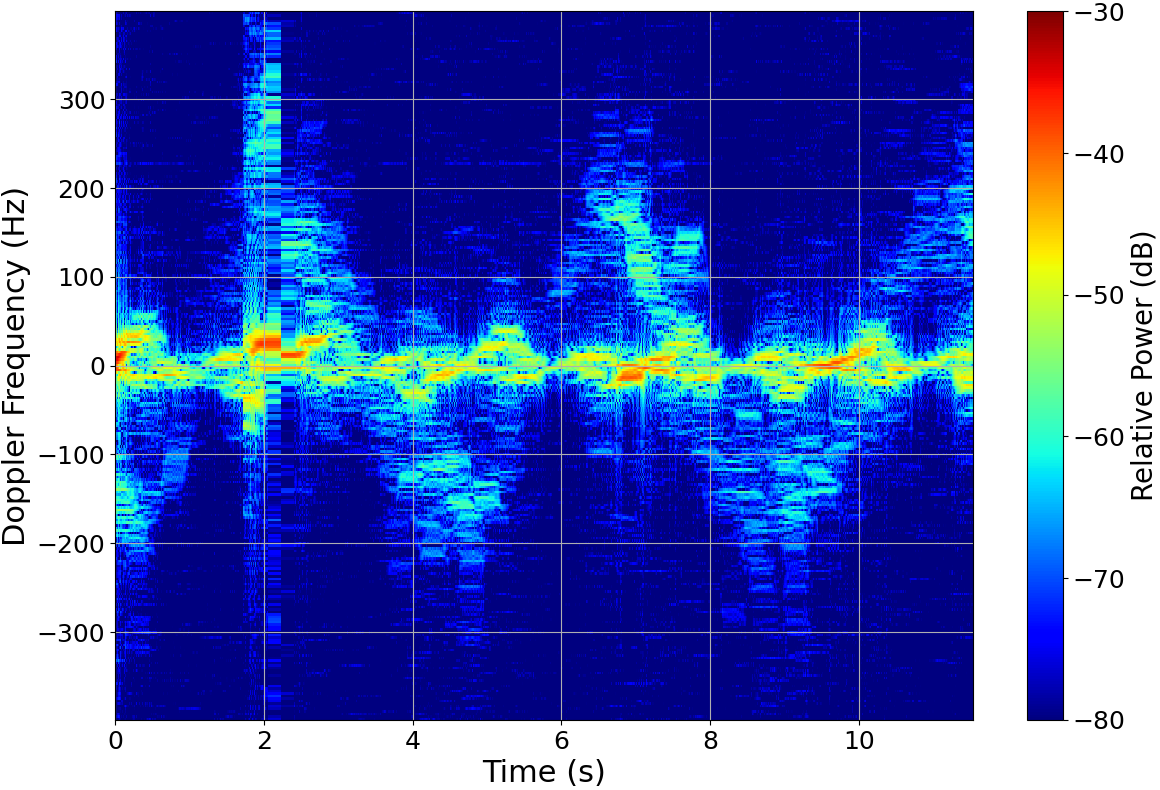}
\caption{Doppler spectrogram under two-person walking scenario.}
\label{fig:multi_person}
\end{figure}

\subsubsection{Multi-Person Walking Scenario}
The system's ability to differentiate superimposed signatures from multiple subjects was evaluated in a two-person walking scenario. Both subjects walked back and forth periodically at different positions. The resulting Doppler spectrogram is presented in Fig.~\ref{fig:multi_person}. It can be observed that there are two distinguishable Doppler patterns. The signature of one human is characterized by a periodic pattern with a main Doppler component of approximately $\pm$200~Hz. Sometimes the frequency reaches $\pm$300~Hz. Meanwhile, there is another high-energy signature concentrated in a lower Doppler range. The frequency is approximately $\pm$50~Hz, corresponding to the other human who walked near the sensing system. These results demonstrate the potential of the system to separate different subjects in a 'blind' sensing scenario.

For a more intuitive analysis, the measured Doppler shifts were converted to their corresponding radial velocities using the standard Doppler formula $v = f_d \lambda / 2$. Here, the wavelength is calculated as $\lambda=c/f_c$, where $c$ is the speed of light and the center frequency $f_c$ is 25.1~GHz, resulting in $\lambda \approx 1.19$~cm. Table~\ref{tab:doppler_summary} summarizes the key Doppler signatures and their corresponding radial velocities from the tested scenarios. From ~\cite{WalkSpeed2016}, it can be found that the mean human walking velocity ranges from 0.94 m/s to 1.43 m/s depending on the age and gender. The detected Doppler shift of 200 Hz corresponds to 1.19 m/s, which is rational by comparing with the normal human walking speed. The successful recognition of various motion patterns demonstrates that the proposed system is promising for real-world ISAC applications in complex environments.

\begin{table}[!t]
\begingroup 
\setlength{\tabcolsep}{6pt} 
\small 
\caption{Summary of Doppler Signatures and Corresponding Radial Velocities}
\label{tab:doppler_summary}
\centering
\begin{tabular}{l c c c}
\Xhline{2\arrayrulewidth}
\textbf{Motion Scenario} & \makecell{\textbf{key Doppler} \\ \textbf{Shift} \\ \textbf{(Hz)}} & \makecell{\textbf{Radial} \\ \textbf{Velocity} \\ \textbf{(m/s)}}  & \makecell{\textbf{Channel} \\ \textbf{Strength} \\ \textbf{(dB)}} \\
\Xhline{\arrayrulewidth}
No Human & N/A & N/A & $-100 \sim -80$ \\
Unidirectional Walk & $+200$ & $+1.19$ & $-55 \sim -20$ \\
Hand Wave & $\pm 350$ & $\pm 2.09$  & $-60 \sim -40$ \\
1-Person Walk (Longit.) & $\pm 200$ & $\pm 1.19$  & $-70 \sim -50$ \\
2-Person Walk (Mixed) & $\pm 200$ & $\pm 1.19$  & $-70 \sim -50$ \\
   & $\pm 50$ & $\pm 0.30$ &  $-55 \sim -40$ \\
\Xhline{2\arrayrulewidth}
\end{tabular}
\endgroup
\end{table}

\subsection{Human Motion Validation}
In order to validate the effectiveness and accuracy proposed sensing system, the Microsoft Azure Kinect DK depth camera was utilized as the ground-truth reference~\cite{MicrosoftAzureKinect}. The Kinect depth camera was deployed to continuously capture the motion of the human subject within its field of view as shown in Fig.~\ref{fig:kinect_setup}. Meanwhile, the proposed sensing system was in operation simultaneously. During the experiment, the subject walked away from the sensing antenna. 

Kinect generates skeleton tracking data with 32 joints~\cite{MicrosoftAzureKinect}. To calculate the velocity of the subject, the coordinates of subject's pelvis joint were chosen. The pelvis was selected since it provides a stable representation of the torso's movement. After extracting the trajectory of pelvis joint, principal component analysis (PCA) was applied to determine the principal motion direction. Then the instantaneous velocity was calculated by each two adjacent recorded points and projected onto the principal motion direction. Finally, the ground-truth reference was obtained as the average velocity by applying a 0.2-second sliding window to suppress high-frequency jitter.
  
 The comparison between the proposed system and Kinect reference is shown in Fig.~\ref{fig:kinect_validation}. The Doppler spectrogram is the sensing results from the proposed system, where the Doppler frequency is converted to velocity. The pink solid line is ground-truth reference from the Kinect. It can be observed that both results show good agreement, which demonstrate the effectiveness and accuracy of the proposed sensing system.

 \begin{figure}[ht]
    \centering
    \includegraphics[width=0.75\linewidth]{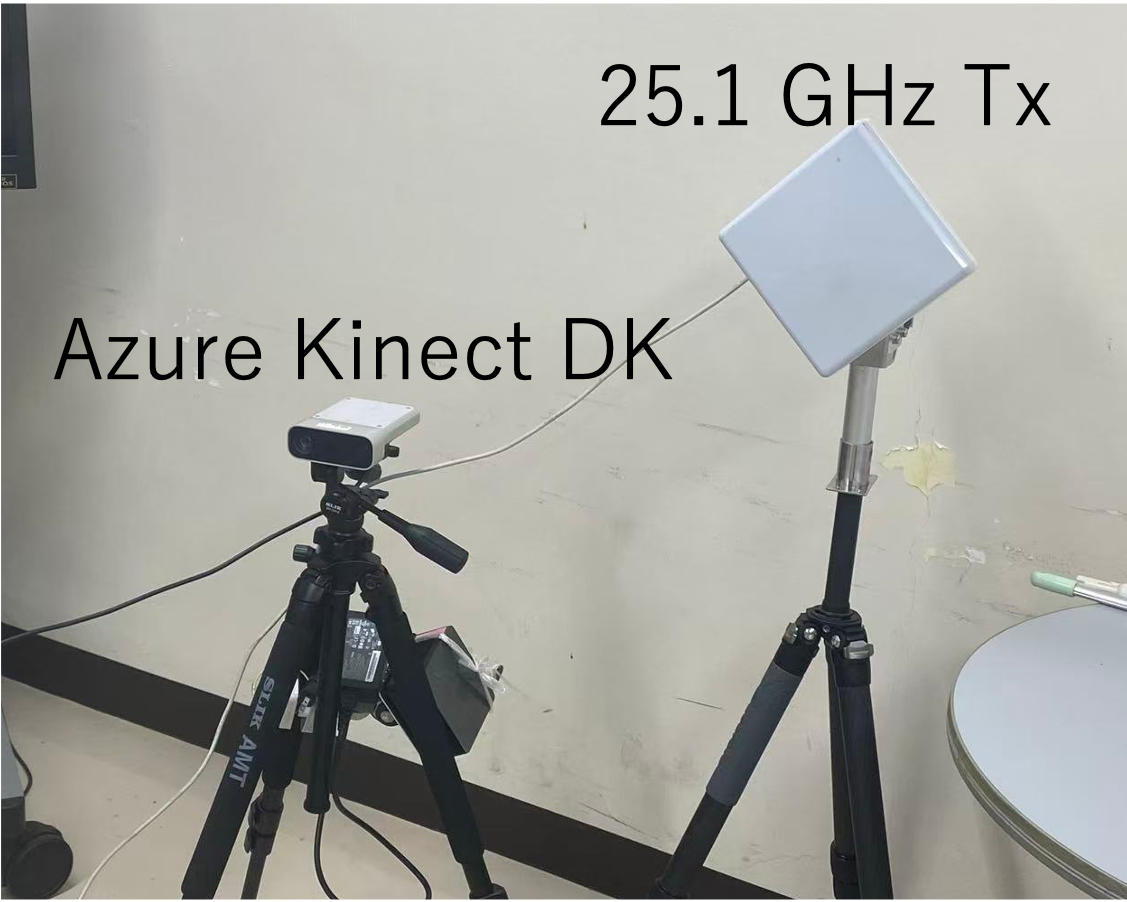}
    \caption{Deployment of Kinect to capture the subject motion}
    \label{fig:kinect_setup}
\end{figure}

\begin{figure}[ht]
    \centering
    \includegraphics[width=0.98\linewidth]{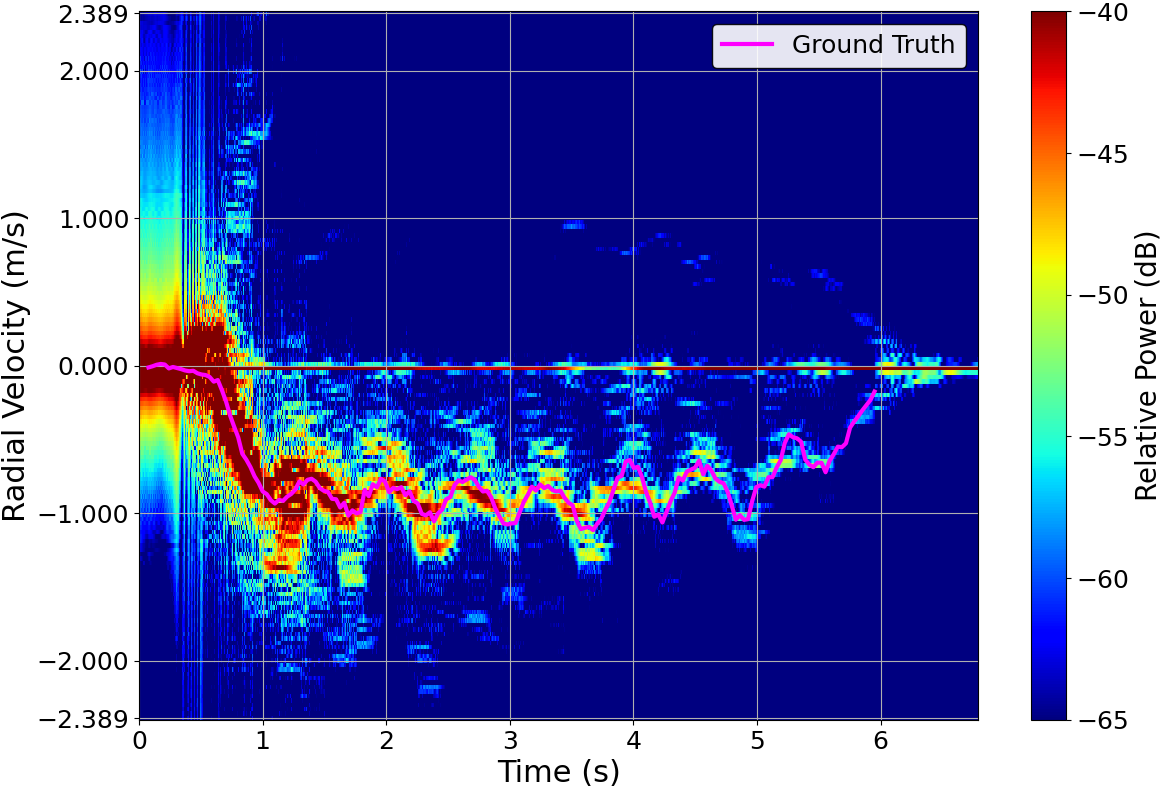}
    \caption{Doppler spectrogram of human walking overlaid with the ground truth velocity (solid line) acquired by the Azure Kinect. }
    \label{fig:kinect_validation}
\end{figure}

\section{Conclusion}
This paper proposed a fully passive mmWave sensing system for blind ISAC based on a COTS SDR. In proposed system, a synchronized dual-receiver differential architecture was utilized which not only successfully circumvents the need for Tx information and active transmissions but also effectively mitigates the common-mode hardware distortions and noises. Furthermore, a signal processing scheme was proposed to realize the sensing with the non-synchronized system.

The accuracy of the proposed system was quantitatively verified in a controlled scenario with metallic plate and validated against ground truth measurements from a depth camera. A time-scaling factor was also introduced to correct the Doppler spectrogram result. In addition, the performance of the system and corresponding signal processing scheme was evaluated in various dynamic scenarios. The measurement results demonstrated that the system is able to capture and differentiate the Doppler patterns of various human motions, including walking, periodic motions, and multiple human motions. It successfully recognized superimposed signatures in the blind sensing context. With the proposed blind ISAC system, it is promising to deploy the mmWave ISAC technology in wider realistic applications such as human activity detection, gait analysis, health monitoring. Challenges remain in the current system, such as estimation of the direction of the object. And the performance may degrade in deep fading conditions. In the future, the system will be extended with array antenna to enable spatial motion analysis. Meanwhile, the experiment and verification in more scenarios will be conducted and different sensing algorithms will be compared.

% \subsection{Limitations}
% The current system has certain limitations, which include:
% \begin{itemize}
%     \item The two-channel hardware restricts comprehensive spatial sensing capabilities.
%     \item Numerical instability can be observed near spectral nulls.
%     \item The NU-STFT computational load requires further optimization.
% \end{itemize}

% \subsection{Future Work}
% Building on this work, future research will focus on addressing the system's current limitations. The primary direction is to advance from motion detection to target localization by upgrading the hardware to a multi-channel receiver platform, such as a 4-channel USRP, to enable AoA estimation. Concurrently, the high computational burden of the NU-STFT will be addressed through algorithmic optimization and GPU acceleration to work towards real-time activity recognition. Finally, the development of more sophisticated algorithms for handling spectral nulls during channel estimation will be pursued to further improve signal fidelity.

\bibliographystyle{IEEEtran}
\bibliography{references}
\end{document}